\documentstyle[prb,aps,eqsecnum,multicol,epsf]{revtex}

\newcommand{\bleq}{\ifpreprintsty
                   \else
                   \end{multicols}\vspace*{-3.5ex}{\tiny
                   \noindent\begin{tabular}[t]{c|}
                   \parbox{0.493\hsize}{~} \\ \hline \end{tabular}}
                   \fi} 
\newcommand{\eleq}{\ifpreprintsty
                  \else
                   {\tiny\hspace*{\fill}\begin{tabular}[t]{|c}\hline
                    \parbox{0.49\hsize}{~} \\
                    \end{tabular}}\vspace*{-2.5ex}\begin{multicols}{2}
                    \fi}
\newcommand{\bcols}{\ifpreprintsty\else\begin{multicols}{2}\fi}
\newcommand{\ecols}{\ifpreprintsty\else\end{multicols}\fi}

\begin{document}
\draft

\title{A strong-coupling expansion for the Hubbard model }

\author{N. Dupuis$^{(1,2)}$ and S. Pairault$^{(2,3)}$  }
\address{(1) Laboratoire de Physique des Solides, Associ\'e au CNRS,
Universit\'e Paris-Sud, 91405 Orsay, France \\ 
(2) Centre de Recherche en Physique du Solide et D\'epartement de Physique, \\
Universit\'e de Sherbrooke, Sherbrooke, Qu\'ebec, Canada J1K 2R1 \\
(3) R\'esidence Palais Soleau, 32 Av. Robert Soleau, 06600 Antibes Juan
Les Pins, France } 
\date{}
\maketitle

\begin{abstract}
We reconsider the strong-coupling expansion for the Hubbard model
recently introduced by Sarker and Pairault {\it et al.} By introducing
slave particles that act as projection operators onto the empty,
singly occupied and doubly occupied atomic states, the perturbation   
theory around the atomic limit distinguishes between processes that do
conserve or do not conserve the total number of doubly occupied
sites. This allows for a systematic $t/U$ expansion that does not break
down at low temperature ($t$ being the intersite  hopping amplitude
and $U$ the local Coulomb repulsion).
The fermionic field becomes a two-component field,
which reflects the presence of the two Hubbard bands. The
single-particle propagator is naturally expressed as a function of a
$2 \times 2$ matrix self-energy. Furthermore, by introducing a time-
and space-fluctuating spin-quantization axis in the functional integral, 
we can expand around a ``non-degenerate'' ground-state where each singly 
occupied site has a well defined spin direction (which may fluctuate
in time). This formalism is used to derive the
effective action of charge carriers in the lower Hubbard band to first order in
$t/U$. We recover the action of the $t$-$J$  model in the spin-hole
coherent-state path integral. We also compare our results with those 
previously obtained by studying fluctuations around the large-$U$
Hartree-Fock saddle point.
\end{abstract}


\bcols

\section{Introduction}

In the last two decades, the
discovery of heavy-fermion compounds, high-$T_c$ superconductors and
organic conductors has revived interest in strongly correlated 
electron systems. Although the first investigations go back to the 1960s, a 
proper understanding of correlation effects in Fermi systems remains a
fundamental issue in condensed-matter physics. Even for the Hubbard model,
\cite{Hubbard63,Gutzwiller63,Kanamori63,Gebhard97}
which is supposed to be one of the simplest (realistic) models
of strongly correlated fermions, exact solutions or well-controlled 
approximations exist only in a few special cases, like in one-dimension
\cite{Voit} or in the limit of infinite dimension.\cite{Georges96} 

In a system where the Coulomb interaction dominates, it is natural to
treat the latter exactly and then consider the kinetic energy within 
perturbation theory. In the Hubbard model, this amounts to 
expanding around the atomic limit (no intersite hopping: $t=0$). [This
kind of expansion will be referred to as {\it strong-coupling}
expansion.] One of the first examples of a strong-coupling expansion
is due to Anderson. Before the introduction of the Hubbard model, he
showed that Fermi systems with a strong local Coulomb repulsion 
are described by an effective Heisenberg Hamiltonian at half-filling
(one particle per site on average). The strong on-site Coulomb
repulsion $U$ completely freezes out the kinetic energy, and virtual
intersite hopping results in an effective antiferromagnetic (AF) exchange 
interaction with coupling constant $J=4t^2/U$.\cite{Anderson59,Bulaevskii66}

The strong-coupling expansion for the Hubbard model presents technical
difficulties that do not show up in standard weak coupling
perturbation theories. Since the atomic Hamiltonian is not quadratic (it
contains the on-site Coulomb repulsion), there is no Wick's theorem so 
that the standard many-body techniques cannot be applied. The perturbation 
theory in the hopping amplitude is based on a cumulant expansion with no 
linked cluster theorem. Given these technical difficulties, it is not 
surprising that contradicting results for the expansion of the free energy and
other thermodynamical quantities can be found in the literature.
\cite{Bulaevskii73,Hone,Beni73,Plischke74,Kubo80,Kubo83,Zhao87,Pan91} 
The calculation of dynamical quantities turns out to be even more
involved. The first attempt to calculate the single-particle Green's function
is due to Hubbard.\cite{Hubbard66} His approach became effective only with the
latter development of a Wick's theorem for Hubbard
operators.\cite{Izyumov92} 
The strong-coupling expansion based on Hubbard operators remains 
however quite cumbersome, since it does not rely on standard many-body
techniques for fermionic or bosonic fields.  

An elegant derivation of the strong-coupling expansion has been
recently introduced by Sarker\cite{Sarker88} and 
Pairault {\it et al.}\cite{Pairault98,Senechal00} Metzner has obtained the 
same results, but his derivation is not as direct.\cite{Metzner91}
The basic idea of this approach is to
decouple the intersite hopping term by means of a Grassmannian 
Hubbard-Stratonovich transformation.\cite{Bourbonnais85} The resulting action
for the  Grassmannian auxiliary field $\psi$ allows for a systematic 
perturbative expansion based on Wick's theorem. The (bare) propagator of the
$\psi$ field being  proportional to the intersite hopping amplitude, a diagram
with $p$ lines (i.e. $p$ propagators) is of order $t^p$. The Green's
functions of the original fermions are simply related to those of the
$\psi$ field and can also be obtained in a systematic way. One of the
difficulties of the strong-coupling perturbation theory is 
that there is now an infinite number of vertices. They are given by
the connected atomic Green's functions and have therefore a
non-trivial time dependence. High-order contributions are not easily
obtained analytically, but can be nevertheless systematically
computed using a specially designed computer program (see 
Ref.~\onlinecite{Pairault98}). 
Another problem lies in the fact that the calculated
Green's functions (to a given order in $t$) do not have the correct analytical
properties. This difficulty was circumvented in
Ref.~\onlinecite{Pairault98} by representing the Green's function as a
finite Jacobi continued fraction, which ensures the correct analytical
behavior.

Besides technical difficulties, the expansion around the atomic limit faces
a more serious problem, namely the existence of two dimensionless expansion 
parameters, $t/U$ and $t/T$ ($T=1/\beta$ is the temperature). Whereas
hopping processes that do change the total 
number of doubly occupied sites are of order $t/U$, there are also processes 
that do not involve the small parameter $t/U$. The former processes
correspond to virtual transitions between the lower (LHB) and upper
(UHB) Hubbard bands, and the latter to intraband propagation. 
[Here we assume that the system is a Mott-Hubbard insulator.]
This implies that expanding around the atomic limit is valid only at
high temperature ($T\gg t$), when intraband hopping processes are
suppressed by the small parameter $t/T$, while we are mainly
interested in the low-temperature regime $T\ll t$. This failure of the
strong-coupling approach was recognized early on, which motivated
a reorganization of the perturbative expansion. Harris and Lange were
the first to derive  an effective
Hamiltonian which describes exactly intraband hopping processes,
while virtual interband transitions are considered perturbatively
to a given order in $t/U$. \cite{Harris67,Chao78,Hirsch85} To first
order in $t/U$, this yields the Hamiltonian of the $t$-$J$ model. \cite{Fulde}

The aim of this paper is to modify the strong-coupling expansion for
the Hubbard 
model introduced in Refs.~\onlinecite{Sarker88,Pairault98} in order to clearly
distinguish between intra and interband processes, thus allowing for an
expansion in $t/U$. Hopping processes that do and those that do not
conserve the 
total number of doubly occupied sites are identified by introducing
slave particles (bosons) which act as
projectors onto the empty, singly occupied, and doubly occupied atomic
states (Sec.~\ref{sec:sce}). \cite{Kotliar86} We emphasize however that the 
slave particles are treated {\it exactly}, which is possible since they 
intervene {\it in fine} only in the atomic limit. The Grassmannian
Hubbard-Stratonovich transformation then requires the introduction of
a two-component fermionic auxiliary field $\psi\equiv (\psi_+,
\psi_-)^T$, where $\psi_+$ ($\psi_-$) corresponds to a particle in the UHB
(LHB). The single-particle propagator is naturally expressed as a
function of a $2\times 2$ matrix self-energy. A simple approximation
for the self-energy in the large-$U$ limit predicts a metal-insulator
transition. We compare this result with the one obtained within the
Hubbard-I approximation. 

In Sec.~\ref{sec:nond}, we further modify our formalism in order to
set up a strong-coupling expansion around a ``non-degenerate''
ground-state where each singly occupied state has a well defined spin
direction (which may fluctuate in time). This is achieved by
introducing a time- and space-fluctuating spin-quantization axis in
the functional integral (Sec.~\ref{subsec:sri}). This provides a
spin-rotation-invariant slave-boson description, which is an
alternative approach to the operator formalism derived by Li {\it et
al.} \cite{Li89} The effective action of charge carriers in the lower Hubbard
band is derived in Sec.~\ref{subsec:eff}. To lowest order in $t/U$, the
coupling between holes and spin fluctuations (which arise from the
dynamics of the spin quantization axis) is described by a U(1) gauge-field
theory, in agreement with previous conclusions.
\cite{Baskaran88,Wiegmann88,Wen89,Lee89,Shankar89,Schulz90,Auerbach91}
To first order in $t/U$, we recover the action of the $t$-$J$
model in the spin-hole coherent-state path integral.\cite{Auerbach91,Auerbach}
We also compare our results with those previously obtained by studying
fluctuations around the large-$U$ Hartree-Fock saddle
point.\cite{Schulz90}

\section{\lowercase{$t$}$/U$ expansion}
\label{sec:sce}

We consider a $D$-dimensional bipartite lattice. 
The Hubbard model is defined by the Hamiltonian 
\begin{equation}
H=-t \sum_{\langle{\bf r},{\bf r}'\rangle,\sigma} (\hat c^\dagger_{{\bf r}
\sigma}\hat c_{{\bf r}'\sigma} +{\rm h.c.} ) + U \sum_{\bf r} \hat n_{{\bf r}
\uparrow} \hat n_{{\bf r}\downarrow} ,
\label{Ham}
\end{equation}
where $\hat c_{{\bf r}\sigma}$ is a fermionic operator for a $\sigma$-spin 
particle at site $\bf r$ ($\sigma=\uparrow,\downarrow$), $\hat n_{{\bf
r}\sigma}=\hat c^\dagger_{{\bf r} \sigma}\hat c_{{\bf r}\sigma}$, and
$\langle {\bf r},{\bf r}'\rangle$ denotes nearest neighbors. 

The hopping term in (\ref{Ham}) does not distinguish between processes
that do or do not conserve the total number of doubly occupied
sites. Such a distinction can be made by writing the fermion operator as
$\hat c_{{\bf r}\sigma}=\hat n_{{\bf r}\bar\sigma}\hat c_{{\bf r}\sigma}
+(1-\hat n_{{\bf r}\bar\sigma}) \hat c_{{\bf r}\sigma}$
[$\bar\sigma=\downarrow(\uparrow)$ for $\sigma=\uparrow(\downarrow)$].
It is more convenient to rewrite this decomposition using the slave
boson representation of the Hubbard model due to Kotliar and
Ruckenstein. \cite{Kotliar86} Introducing two fermionic ($\hat
f_{{\bf r}\uparrow},\hat f_{{\bf r}\downarrow}$) and four bosonic
($\hat e_{\bf r},\hat p_{{\bf r}\uparrow},\hat p_{{\bf r}\downarrow},\hat
d_{\bf r}$) operators,  the four atomic states (empty, singly
occupied, and doubly occupied) are expressed as 
\begin{eqnarray}
|0,{\bf r}\rangle &=& \hat e^\dagger_{\bf r} |{\rm vac}\rangle , \nonumber \\
|\sigma,{\bf r}\rangle &=& \hat c^\dagger_{{\bf r}\sigma} |0,{\bf r}\rangle =
\hat p^\dagger_{{\bf r}\sigma} 
\hat f^\dagger_{{\bf r}\sigma} |{\rm vac}\rangle \,\,\,\,\,\,
(\sigma=\uparrow,\downarrow) , \nonumber \\ 
|\uparrow\downarrow,{\bf r}\rangle &=& 
\hat c^\dagger_{{\bf r}\uparrow} \hat c^\dagger_{{\bf r}\downarrow} 
|0,{\bf r}\rangle = \hat d^\dagger_{\bf r} 
\hat f^\dagger_{{\bf r}\uparrow} \hat f^\dagger_{{\bf r}\downarrow} 
|{\rm vac}\rangle ,
\label{atst}
\end{eqnarray}
where $|{\rm vac}\rangle$ denotes the vacuum of the enlarged Hilbert
space. This enlarged space contains unphysical states which can be
eliminated by imposing the set of constraints\cite{Kotliar86}
\begin{eqnarray}
Q^{(1)}_{\bf r} &=& \hat e^\dagger_{\bf r} \hat e_{\bf r} + \sum_\sigma \hat
p^\dagger_{{\bf r}\sigma} \hat p_{{\bf r}\sigma} +\hat d^\dagger_{\bf
r} \hat d_{\bf r} -1 = 0 , \nonumber \\
Q^{(2)}_{{\bf r}\sigma} &=& \hat f^\dagger_{{\bf r}\sigma} \hat
f_{{\bf r}\sigma} - \hat p^\dagger_{{\bf r}\sigma} \hat p_{{\bf r}\sigma}
-\hat d^\dagger_{\bf r} \hat d_{\bf r} = 0 \,\,\,\,\,\,
(\sigma=\uparrow,\downarrow) ,
\label{con}
\end{eqnarray}
at each lattice site $\bf r$. The bosonic operators $\hat e_{\bf r}$,
$\hat p_{{\bf r}\sigma}$ and $\hat d_{\bf r}$ act respectively as
projection operators 
onto the empty, singly occupied (with spin $\sigma$) and doubly
occupied states. The Hubbard Hamiltonian (\ref{Ham}) is
rewritten as
\begin{equation}
H = - t \sum_{\langle{\bf r},{\bf r}'\rangle,\sigma} \sum_{\alpha,\alpha'=\pm}
(\hat\gamma^\dagger_{{\bf r}\alpha\sigma} \hat\gamma_{{\bf r}'\alpha'\sigma}
+ {\rm h.c.} ) 
+ U \sum_{\bf r}\hat d^\dagger_{\bf r}\hat d_{\bf r} ,
\label{Ham1}
\end{equation}
where we have introduced the operators 
\begin{eqnarray}
\hat\gamma_{{\bf r}-\sigma} &=& \hat e^\dagger_{\bf r}\hat p_{{\bf
r}\sigma} \hat f_{{\bf r}\sigma} , \nonumber \\ 
\hat\gamma_{{\bf r}+\sigma} &=& \hat d_{\bf r}\hat p^\dagger_{{\bf
r}\bar\sigma} \hat f_{{\bf r}\sigma} ,
\label{gam}
\end{eqnarray}
which annihilate a particle in the LHB and the UHB, respectively. 
The advantage of the form (\ref{Ham1}) of the Hamiltonian is that the
hopping term now distinguishes between intraband ($\alpha=\alpha'$)
and interband ($\alpha=-\alpha'$) processes. 

The partition function $Z={\rm Tr}\,e^{-\beta(H-\mu N)}$ can be
written as a functional integral over fermionic ($f$) and bosonic
($e,p,d$) fields ($N=\sum_{{\bf r}\sigma}\hat f^\dagger_{{\bf
r}\sigma} \hat f_{{\bf r}\sigma}$ is the total number of particles and
$\mu$ the chemical potential)\cite{Kotliar86}
\begin{eqnarray}
Z &=& \int d\lambda \int {\cal D}[f,e,p,d] \exp \Bigl \lbrace 
- S_{\rm at}[c,e,p,d;\lambda] \nonumber \\ && 
+ \sum_{{\bf r},{\bf r}',\alpha,\alpha',\sigma} \int d\tau \,
\gamma^*_{{\bf r}\alpha\sigma} t_{{\bf rr}'} \gamma_{{\bf
r}'\alpha'\sigma} \Bigr \rbrace ,
\label{Z1}
\end{eqnarray}
where $\tau$ is an imaginary time varying between 0 and $\beta$.
$t_{{\bf rr}'}$ equals $t$ if $\bf r$ and ${\bf r'}$ are first 
neighbors and vanishes otherwise. The variables $\gamma_{{\bf
r}\alpha\sigma}$ are defined from (\ref{gam}) by replacing the
operators by fermionic or bosonic fields. $\lambda\equiv
(\lambda^{(1)}_{\bf r},\lambda^{(2)}_{{\bf r}\sigma})$ denotes a set
of time-independent Lagrange multipliers that impose the constraints
(\ref{con}). $S_{\rm at}$ is the atomic part of the action. \cite{Kotliar86}
 
We are now in a position to apply the decoupling procedure introduced
in Refs.~\onlinecite{Sarker88,Pairault98}. Since the field
$\gamma_{{\bf r}\alpha\sigma}$ appearing in the hopping term [Eq.~(\ref{Z1})]
carries a band index ($\alpha=\pm$), the auxiliary fermionic field
involved in the Grassmannian Hubbard-Stratonovich transformation is a
two-component field
\begin{equation}
\psi_{{\bf r}\sigma} = {\psi_{{\bf r}+\sigma} \choose \psi_{{\bf
r}-\sigma}}. 
\end{equation}
This is the main modification with respect to the strong-coupling
expansion of Refs.\onlinecite{Sarker88,Pairault98}. The procedure to derive 
the action of the auxiliary field is similar and is
briefly described below. The partition function reads
\begin{eqnarray}
Z &=& \int {\cal D}[\psi] \exp \Bigl \lbrace -\sum_{a,b} \psi^*_a\hat
t_{ab}^{-1} \psi_b \Bigr \rbrace \nonumber \\ && \times
\int d\lambda \int {\cal D}[f,e,p,d] \exp \Bigl \lbrace 
- S_{\rm at} +\sum_a(\psi^*_a \gamma_a + {\rm c.c.}) \Bigr \rbrace ,
\nonumber \\ && 
\end{eqnarray}
where we use the notation
\begin{equation}
\psi_a \equiv \psi_{{\bf r}_a\alpha_a\sigma_a}(\tau_a), \,\,\,\,\,\, 
\sum_a \equiv \sum_{{\bf r}_a,\alpha_a,\sigma_a} \int d\tau_a.
\end{equation}
We denote by $\hat t_{{\bf rr}'}$ the hopping matrix to emphasize that
it acts on the band index of the $\psi$ field. It can be written as
the tensor product of $t_{{\bf rr}'}$ with a $2\times 2$ matrix acting
on the band indices: 
\begin{equation}
\hat t_{{\bf rr}'} = t_{{\bf rr}'} \otimes 
\left( \begin{array}{cc} 1 & 1 \\ 1 & 1 \end{array} \right). 
\end{equation}
Note that the matrix $\hat t$ cannot be inverted. This is not a real
difficulty since the final results can always be expressed in terms of
the matrix $\hat t$, the inverse matrix $\hat t^{-1}$ appearing only
at intermediate stages in the calculation. One could also consider a
more general matrix where interband and intraband hopping processes
have different amplitudes, which ensures the existence of the inverse
matrix $\hat t^{-1}$.

Performing the functional integration over the fields $f,e,p,d$ and
the Lagrange multipliers, we obtain
\begin{eqnarray}
\int d\lambda \int {\cal D}[f,e,p,d] \exp \Bigl \lbrace 
- S_{\rm at} +\sum_a(\psi^*_a \gamma_a + {\rm c.c.}) \Bigr \rbrace &&
\nonumber \\ 
= Z_{\rm at} \exp \Bigl \lbrace W[\psi^*,\psi] \Bigr \rbrace , && 
\label{W1}
\end{eqnarray}
where $Z_{\rm at}$ is the partition function in the atomic limit
($t=0$) and $W[\psi^*,\psi]$ the generating functional of the connected atomic
Green's functions
\begin{eqnarray}
G^{Rc}_{\lbrace a_i,b_i\rbrace} &=&  (-1)^R \langle
\gamma_{a_1}\cdots\gamma_{a_R}\gamma_{b_R}^* \cdots \gamma_{b_1}^*
\rangle _{\rm at,c} \nonumber \\  
&=& \frac{\delta^{(2R)} W[\psi^*,\psi]}{\delta\psi^*_{a_1}\cdots
\delta \psi^*_{a_R}\delta \psi_{b_R}\cdots \delta \psi_{b_1}} 
\Biggr |_{\psi^*=\psi=0} .
\label{GR}
\end{eqnarray}
Thus the action of the $\psi$ field is given by 
\begin{equation}
S[\psi^*,\psi]=\sum_{a,b} \psi^*_a \hat t^{-1}_{ab} \psi_b -
W[\psi^*,\psi],
\label{action1}
\end{equation}
where $W[\psi^*,\psi]$ can be obtained explicitly by inverting
Eq.~(\ref{GR}):
\begin{equation}
W[\psi^*,\psi] = \sum_{R=1}^\infty \frac{(-1)^R}{(R!)^2}
\sum_{a_i,b_i}' \psi^*_{a_1}\cdots \psi^*_{a_R} \psi_{b_R}\cdots
\psi_{b_1} G^{Rc}_{\lbrace a_i,b_i\rbrace} .
\label{W2}
\end{equation}
The primed summation in (\ref{W2}) reminds us that all the fields in a
given product $\psi^*_{a_1}\cdots \psi_{b_1}$ share the same value of
the site index. Note that the integration of the bosonic fields
and the Lagrange multipliers has been done exactly. 
A difference with the strong-coupling expansion introduced in 
Refs.~\onlinecite{Sarker88,Pairault98} is that the effective action of
the auxiliary field $\psi$ involves the connected atomic
Green's functions {\it projected} onto empty, singly occupied, or doubly
occupied states. For instance, for the single-particle Green's
function, we find (see Appendix \ref{appI})
\begin{eqnarray}
G_{-\sigma}(i\omega)&=&\frac{1-n_{\bar\sigma}}{i\omega+\mu}, \nonumber \\
G_{+\sigma}(i\omega)&=&\frac{n_{\bar\sigma}}{i\omega+\mu-U},
\end{eqnarray}
for the LHB and UHB atomic Green's functions, respectively. 
Here $\omega$ is a fermionic Matsubara frequency, and $n_\sigma$ the
average number of $\sigma$-spin particles per site.
The atomic Green's function of the original fermions ($c$) is
simply given by the sum of $G_{-\sigma}$ and $G_{+\sigma}$.

The free propagator of the auxiliary field is $-\hat t$,
i.e. $-\langle \psi_a\psi^*_b\rangle_{\rm free}=-\hat t_{ab}$, and we may
now use Wick's theorem to treat the interaction terms contained in
$W[\psi^*,\psi]$ perturbatively. A diagram with $p$ lines (i.e. $p$
free propagators) is of order $t^p$. The matrix structure of $\hat t$
allows to distinguish between processes of order $t/U$ and those
of order $t/T$. The former correspond to off-diagonal elements of
$\hat t_{{\bf rr}'}$ and the latter to diagonal elements. 

Although the action (\ref{action1}) allows to set up a systematic
$t/U$ expansion, some difficulties remain. The first step would be to
find an exact or approximate solution to lowest order in $t/U$, before
considering interband coupling perturbatively. But this requires to
consider an infinite number of vertices, with no obvious small
expansion parameter since $t/T\gg 1$ at low temperature. Except in one
dimension, where an exact solution exists, \cite{Gebhard97} this
problem is still unsolved. In
Sec.~\ref{sec:nond}, we propose a slightly different formulation which
solves (at least partially) this problem. 
Nevertheless, it is interesting to discuss approximate
expressions of the single-particle Green's function that can be obtained from
Eq.~(\ref{action1}). This is done in the next section.

\subsection*{Single-particle Green's function}

The single-particle Green's function $\hat {\cal G}_{ab}=-\langle
\gamma_a\gamma^*_b\rangle$ is related to the propagator of the
auxiliary field $\hat {\cal T}_{ab}=-\langle \psi^*_a\psi_b\rangle$ by
(in matrix form)\cite{Pairault98}
\begin{equation}
\hat{\cal G}=\hat t^{-1} + \hat t^{-1} \hat{\cal T} \hat t^{-1}.
\end{equation}
Introducing the self-energy $\hat\Gamma$ of the auxiliary field, $\hat
{\cal T}^{-1}=-\hat t^{-1}-\hat\Gamma$, we obtain
\begin{equation}
\hat{\cal G}^{-1} = \hat t + \hat\Gamma^{-1}. 
\label{hatG1}
\end{equation}
The self-energy $\hat\Gamma$ is now a matrix with respect to
the band indices. In reciproqual space, $\hat\Gamma_\sigma({\bf k},i\omega)$
is a $2\times 2$ matrix. Consider a $R$-particle vertex for the field
$\psi$ [Eqs.~(\ref{action1}) and (\ref{W2})]. If $p$ ($p\leq R$) incoming
particles are in the LHB, then there are exactly $p$ outgoing
particles in the LHB. One can then easily convince oneself that 
the $2\times 2$ matrix
self-energy $\hat \Gamma_{\sigma}({\bf k},i\omega)$ is diagonal (even
though the hopping matrix $\hat t$ has off-diagonal elements with
respect to band indices). We denote its two components by
$\Gamma_{-\sigma}({\bf k},i\omega)$ and $\Gamma_{+\sigma}({\bf k},i\omega)$. 

The Green's function of the original fermions is related to $\hat
{\cal G}$ by 
\begin{equation}
{\cal G}_\sigma({\bf k},i\omega) = \sum_{\alpha,\alpha'} 
\hat {\cal G}_{\alpha\alpha',\sigma} ({\bf k},i\omega) .
\end{equation}
From Eq.~(\ref{hatG1}), we obtain 
\begin{equation}
{\cal G}_\sigma({\bf k},i\omega) = \frac{\Gamma_\sigma^{\rm tot}({\bf
k},i\omega)}{1+t_{\bf k}\Gamma_\sigma^{\rm tot}({\bf k},i\omega)},
\label{calG3}
\end{equation}
where 
\begin{equation}
\Gamma_\sigma^{\rm tot}({\bf k},i\omega)= \sum_\alpha
\Gamma_{\alpha\sigma}({\bf k},i\omega),
\label{G1}
\end{equation}
and $t_{\bf k}$ is the Fourier transform of $t_{{\bf rr}'}$. [In the
absence of interaction, the band energy is $\epsilon_{\bf k}=-t_{\bf
k}$.]  Eq.~(\ref{G1}) is nothing but the result obtained in
Ref.~\onlinecite{Pairault98}, where 
$\Gamma^{\rm tot}$ was calculated in perturbation theory with respect
to the hopping amplitude $t$. To the extent where the self-energy
$\hat\Gamma$ can be calculated exactly, our strong-coupling expansion
is similar to that of Ref.~\onlinecite{Pairault98}. 

\subsubsection*{The large-$U$ limit}

Interband transitions are suppressed when $U\to\infty$, so that the
off-diagonal elements of $\hat t_{\bf k}$ do not play any role. In the
large-$U$ limit, it is natural to make a diagonal approximation where
interband transitions are neglected. The Green's function is then determined by
\begin{equation}
\hat {\cal G}^{-1}_\sigma({\bf k},i\omega) =  
\left( \begin{array}{cc}  t_{\bf k}+\Gamma_{+\sigma}({\bf k},i\omega)^{-1}
& 0 \\ 0 & t_{\bf k}+\Gamma_{-\sigma}({\bf k},i\omega)^{-1} \end{array}
\right)  .
\label{calG4}
\end{equation}
Eq.~(\ref{calG4}) gives
\begin{equation}
{\cal G}_\sigma({\bf k},i\omega) = \frac{1}{t_{\bf k}+\Gamma_{+\sigma}({\bf
k},i\omega)^{-1}} + \frac{1}{t_{\bf k}+\Gamma_{-\sigma}({\bf 
k},i\omega)^{-1}}. 
\label{calG1}
\end{equation}
The calculation of $\Gamma_{+\sigma}$ and $\Gamma_{-\sigma}$
remains a very difficult task. The simplest approximation (lowest
order in $t$) consists in retaining only the Gaussian part of the action
$S[\psi^*,\psi]$ [Eq.~(\ref{action1})]: 
\begin{equation}
\Gamma_{\alpha\sigma}({\bf k},i\omega) = G_{\alpha\sigma}(i\omega).
\label{Gam1}
\end{equation}
From (\ref{calG1}) we deduce  
\begin{equation}
{\cal G}_\sigma({\bf k},i\omega) =
\frac{1-n_{\bar\sigma}}{i\omega+\mu+(1-n_{\bar\sigma})t_{\bf k}} + 
\frac{n_{\bar\sigma}}{i\omega+\mu-U+n_{\bar\sigma}t_{\bf k}} .
\label{calG2}
\end{equation}
Retaining only the Gaussian part of the action $S[\psi^*,\psi]$
[Eq.~(\ref{Gam1})] is clearly 
unjustified, since this amounts to ignoring terms of order $(t/T)^p$
($p\geq 1$) which are not small at low temperature. Nevertheless, this
approximation yields interesting results. At half-filling ($\mu=U/2$
and $n_\sigma=1/2$), we obtain two Hubbard bands with dispersions
\begin{eqnarray}
E_{+\sigma}({\bf k}) &=& \frac{U-t_{\bf k}}{2} , \nonumber \\ 
E_{-\sigma}({\bf k}) &=& -\frac{U+t_{\bf k}}{2} .
\end{eqnarray}
The Hubbard bands can be seen as due to the broadening of the atomic
levels when the intersite hopping is switched on. They are separated
by the Mott-Hubbard gap $\Delta=U-B/2$ ($B$ is the total bandwidth in
the absence of interaction), which closes for the critical value
$U_c=B/2$ of the Coulomb interaction. This simple approximation
[Eq.~(\ref{calG2})] therefore predicts a metal-insulator transition at
half-filling. It should be noted however that the bandwidth of the
Hubbard bands is only half the total bandwidth
$B$ in the absence of interaction. This is clearly an artifact of our
approximation where only the particle (or the hole) that was added to
the system can propagate. If the particle injected in the UHB has spin
$\sigma$, it can hop to the neighboring site only if the latter is
occupied by a $\bar\sigma$-spin particle, so that the hopping
amplitude is reduced by a factor $n_{\bar\sigma}$. [A similar results holds 
for the propagation of a hole in the LHB.] We expect the exact bandwidth
of the Hubbard bands 
to be of order $B$ in agreement with Mott's arguments.\cite{Mott} 
For the same reason, we find that the Mott gap $\Delta$
separating the two bands closes for the critical value $U_c=B/2$,
while Mott suggested $U_c\sim B$. 

The fact that the single-particle propagator is naturally expressed as
a function of a $2\times 2$ matrix self-energy has also been noted by
Logan and Nozi\`eres. \cite{Logan} This result is reported
in Ref.~\onlinecite{Gebhard97} where an expression similar to
Eq.~(\ref{calG2}) is discussed [Eq.~(3.38) of
Ref.~\onlinecite{Gebhard97}]. The difference with
Eq.~(\ref{calG2}) is that the Hubbard bands have the full bandwidth
$B$. 

The result obtained in this section should be compared with the Hubbard-I
approximation,\cite{Hubbard63} which predicts the opening of a charge
gap for any finite value of the Coulomb repulsion $U$. The Hubbard-I
approximation is obtained from Eq.~(\ref{calG3}) with the lowest-order
contribution in $t$ for the self-energy $\Gamma^{\rm tot}$,
i.e. $\Gamma^{\rm tot}_\sigma=\sum_\alpha
G_{\alpha\sigma}$. \cite{Pairault98} It can also be obtained
from Eqs.~(\ref{calG4}) and (\ref{Gam1}) by including in (\ref{calG4}) the
off-diagonal elements of the matrix $\hat t_{\bf k}$. Thus we see that
the Hubbard-I approximation includes interband coupling without
giving a correct description of the Hubbard bands to zeroth order in
$t/U$.

\section{Non-degenerate \lowercase{$t$}$/U$ expansion}
\label{sec:nond}

An important feature of the strong-coupling perturbative theory is
that one expands around a state which has a huge degeneracy near
half-filling, since a particle on a singly-occupied site can have its
spin up or down. In this section we propose an alternative
approach which allows to expand around a ``non-degenerate''
ground-state. 

In our formalism, the degeneracy of the ground-state shows up in the
atomic Green's functions $G^{Rc}$ which contain all the information
about the atomic limit ($t=0$). These Green's functions have to be
calculated with the chemical potential $\mu$ defined by the {\it full}
Hubbard model. At half-filling, $\mu=U/2$ due to particle-hole
symmetry. In the presence of a small concentration of holes,
the chemical potential lies near the top of the LHB, i.e. above
the energy of the isolated atomic state. Therefore, at low enough
temperature (when $T$ is much smaller than  half the bandwidth of the
LHB), the system in the atomic limit is in its ground-state with
exactly one particle per site (even in the presence of a finite
concentration of holes). To suppress the degeneracy of the
ground-state, we then proceed as follows. We impose the particles to
have spin up. In order to restore spin-rotation invariance, we then
allow the spin-quantization axis to fluctuate in time and space, and
integrate over all possible configurations in the functional
integral. For a
given configuration of the spin-quantization axis, the expansion in
the hopping amplitude $t$ becomes a non-degenerate perturbation
theory. We shall 
show that this formulation allows to derive the effective action of
charge carriers in the LHB to first order in
$t/U$. We recover the action of the $t$-$J$ model in the spin-hole
coherent-state path integral. \cite{Auerbach91,Auerbach}
We also comment on the effective action
obtained by studying fluctuations around the large-$U$
Hartree-Fock saddle point.\cite{Schulz90} Within our formalism, the
results of this approach 
are reproduced by retaining only the quadratic part of the
auxiliary-field action. This Gaussian approximation however does not
capture all processes of order $t/U$, and therefore does not allow to
derive the $t$-$J$ model. 

In section \ref{subsec:sri}, we express the partition function as a functional
integral where the spin-quantization axis fluctuates in time and
space. This provides a spin-rotation-invariant slave-boson approach to
the Hubbard model. An alternative approach, based on the operator
formalism, has been proposed by Li {\it et al.}\cite{Li89} 
The effective action of charge carriers in the LHB is then derived in
Sec.~\ref{subsec:eff}.

\subsection{Spin-rotation-invariant slave-boson approach}
\label{subsec:sri}

The procedure to introduce a fluctuating spin-quantization axis in the
functional integral has been given by Schulz. \cite{Schulz90,Weng91}
It simply amounts to introducing a new field related to the
old one by a unitary matrix $R_{\bf r}$, which rotates the
spin-quantization axis  at site $\bf r$ and time $\tau$ from $\hat{\bf z}$
to a new axis defined by the unit vector ${\bf\Omega}_{\bf r}$
($\hat{\bf z}$ is the unit vector along the $z$ axis). 
In the present case, the procedure is slightly more complicated due to the
presence of the slave bosons. Indeed it has been shown in
Ref.~\onlinecite{Li89} that the bosonic field $p_{\bf r}$ transforms under a
spin rotation as a $2\times 2$ spin matrix field $p_{{\bf
r}\sigma\sigma'}$ rather than as a two-component spinor field $p_{{\bf
r}\sigma}$. We show below that the introduction of a $2\times 2$ spin
matrix field can be avoided if one builds the functional integral from
the very beginning.

Let us first go back to the operator formalism. At each lattice site,
we rotate the local spin-quantization axis by introducing a new
fermionic operator
\begin{equation}
\hat \phi_{\bf r}=R^\dagger_{\bf r}\hat c_{\bf r},
\end{equation}
where $\hat c_{\bf r}=(\hat c_{{\bf r}\uparrow},\hat c_{{\bf
r}\downarrow})^T$, $\hat\phi_{\bf r}=(\hat\phi_{{\bf r}\uparrow},\hat\phi_{{\bf
r}\downarrow})^T$, and $R_{\bf r}$ is a unitary SU(2)/U(1) rotation matrix
defined by
\begin{equation}
R_{\bf r}\sigma_zR^\dagger_{\bf r} =\bbox{\sigma}\cdot {\bf
\Omega}_{\bf r}. 
\end{equation}
$\bbox{\sigma}=(\sigma_x,\sigma_y,\sigma_z)$ stands for the Pauli
matrices. The field $\hat\phi_{\bf r}$ has its spin-quantization axis
along the (time-independent) unit vector ${\bf \Omega}_{\bf r}$. The
U(1) gauge freedom is due to rotations around the $z$ axis that do not
change the state of the system. A convenient gauge choice is 
\begin{equation}
R_{\bf r} = \left (
\begin{array}{cc}
\cos(\theta_{\bf r}/2) & -e^{-i\varphi_{\bf r}}\sin(\theta_{\bf r}/2) \\
e^{i\varphi_{\bf r}}\sin(\theta_{\bf r}/2) & \cos(\theta_{\bf r}/2) 
\end{array}
\right ) ,
\label{Rdef}
\end{equation}
where $\theta_{\bf r},\varphi_{\bf r}$ are the usual polar angles
determining the direction of ${\bf\Omega}_{\bf r}$. 

The interaction term being rotation invariant, the Hamiltonian is rewritten as
\begin{equation}
H=-\sum_{{\bf r},{\bf r}'} \hat\phi^\dagger_{\bf r} R^\dagger_{\bf
r}t_{{\bf rr}'}R_{{\bf r}'} \hat\phi_{{\bf 
r}'} + U \sum_{\bf r} \hat\phi^\dagger_{{\bf r}\uparrow}
\hat\phi_{{\bf r}\uparrow}  \hat\phi^\dagger_{{\bf r}\downarrow}
\hat\phi_{{\bf r}\downarrow} .
\end{equation}
As in Sec.~\ref{sec:sce}, we introduce fermionic ($\hat
f_{{\bf r}\sigma}$) and bosonic ($\hat e_{\bf r},\hat p_{{\bf
r}\sigma},\hat d_{\bf r}$) operators
at each lattice site (with spin-quantization axis given by ${\bf
\Omega}_{\bf r}$) which satisfy the constraints (\ref{con}). In this
slave boson representation, the Hamiltonian is given by
\begin{equation}
H=-\sum_{{\bf r},{\bf r}',\alpha,\alpha'} \hat\gamma^\dagger_{{\bf
r}\alpha} R^\dagger_{\bf r}t_{{\bf rr}'}R_{{\bf r}'} \hat\gamma_{{\bf
r}'\alpha'} + U\sum_{\bf r} \hat d^\dagger_{\bf r}\hat d_{\bf r}, 
\end{equation}
where $\hat\gamma_{{\bf r}\alpha}=(\hat\gamma_{{\bf r}\alpha\uparrow}, 
\hat\gamma_{{\bf r}\alpha\downarrow})^T$. $\hat\gamma_{{\bf
r}\alpha\sigma}$ is defined by Eq.~(\ref{gam}). 

In order to express the partition function as a
functional integral, we divide the ``time'' interval $\beta$ into $M$ steps:
\begin{equation}
Z= {\rm Tr} \Bigl ( e^{-\epsilon (H -\mu N)}
\cdots  e^{-\epsilon (H -\mu N)} \Bigr ) ,
\label{Z2bis}
\end{equation}
where $\epsilon=\beta/M$. We then introduce $M-1$ times the closure
relation using mixed fermion-boson coherent states. An important
point here is that the local spin-quantization axis ${\bf \Omega}_{\bf r}$
depends a priori on the discrete ``time'' $\tau_k=k\beta/M$ ($k=0,\cdots
M$) and therefore becomes a dynamical variable ${\bf \Omega}_{\bf r}(\tau)$
in the continuum time limit ($M\to \infty$). Spin-rotation invariance
is obtained by summing over all possible configurations of the unit
vector field ${\bf \Omega}_{\bf r}$, i.e. 
\begin{equation}
Z \to \int \prod_{\bf r} \prod_{k=1}^M \frac{d{\bf \Omega}_{\bf
r}(\tau_k)}{4\pi} Z \equiv \int {\cal D}{\bf \Omega}\, Z.
\end{equation}
The details of this derivation are given in Appendix \ref{appII}. We obtain
\begin{equation}
Z= \int {\cal D}{\bf \Omega}\int d\lambda \int {\cal D}[f,e,p,d]\, e^{-S},
\label{Z3}
\end{equation}
where $\lambda\equiv (\lambda^{(1)}_{\bf r},\lambda^{(2)}_{{\bf
r}\sigma})$ denotes a set of time-independent Lagrange multipliers
that impose the constraints (\ref{con}). The action is given by
\begin{equation}
S = S_{\rm at} - \sum_{{\bf r},{\bf r}',\alpha,\alpha'} \int d\tau \,
\gamma^\dagger_{{\bf r}\alpha} R^\dagger_{\bf r}  
t_{{\bf rr}'} R_{{\bf r}'} \gamma_{{\bf r}'\alpha'} .
\label{Stot}
\end{equation}
The atomic part reads 
\begin{equation}
S_{\rm at}= S^{(0)}_{\rm at} + \int d\tau\, \sum_{{\bf r},\sigma,\sigma'}
f^*_{{\bf r}\sigma} p^*_{{\bf r}\sigma} (R^\dagger_{\bf r}\dot
R_{\bf r})_{\sigma\sigma'} f_{{\bf r}\sigma'} p_{{\bf r}\sigma'},
\label{Sat1}
\end{equation}
where $\dot R_{\bf r}=\partial_\tau R_{\bf r}$. $S^{(0)}_{\rm at}$ is
the atomic action for a time-independent spin-quantization
axis [see Eq.~(\ref{Sat0})]. Fluctuations of the spin-quantization axis
modify the hopping 
term [Eq.~(\ref{Stot})] and induce an additional term in the atomic
action [Eq.~(\ref{Sat1})]. As shown below, the latter is related to
the Berry phase term of a singly occupied site (which behaves as a
spin).

\subsection{Effective action of charge carriers in the LHB}
\label{subsec:eff}

The partition function (\ref{Z3}) obviously preserves spin-rotation
invariance. As discussed above, we can now assume that the LHB is
populated only 
with up-spin particles (in the local spin reference frame defined by
${\bf\Omega}_{\bf r}$).\cite{note1} This implies that only down-spin
particles can be introduced in the UHB. Spin-rotation invariance is
maintained since we integrate over all configurations of the unit vector field
${\bf\Omega}_{\bf r}$ in the functional integral.

The assumption that the LHB is populated only with up-spin particles
is imposed by eliminating the $p_\downarrow$ field.  The
action then reads 
\bleq
\begin{eqnarray}
S&=&S_{\rm at}-\sum_{{\bf r},{\bf r}'} \int d\tau\, \gamma^\dagger_{\bf r}
R^\dagger_{\bf r}t_{{\bf rr}'}R_{{\bf r}'} \gamma_{{\bf r}'},
\nonumber \\ 
S_{\rm at}&=& S^{(0)}_{\rm at} + \int d\tau\, \sum_{\bf r}
f^*_{{\bf r}\uparrow} p^*_{{\bf r}\uparrow}A^0_{{\bf r}z}  f_{{\bf
r}\uparrow} p_{{\bf r}\uparrow}, 
\nonumber \\ 
S_{\rm at}^{(0)} &=& \sum_{\bf r}
\int d\tau\, \Bigl [ -i\lambda^{(1)}_{\bf r}+\sum_\sigma  f^*_{{\bf r}\sigma}
(\partial_\tau-\mu+i\lambda^{(2)}_{{\bf r}\sigma}) f_{{\bf r}\sigma} + 
e^*_{\bf r} (\partial_\tau+i\lambda^{(1)}_{\bf r}) e_{\bf r}
\nonumber \\ && + p^*_{{\bf r}\uparrow}
(\partial_\tau+i\lambda^{(1)}_{\bf r}-i\lambda^{(2)}_{{\bf r}\uparrow})
p_{{\bf r}\uparrow}
+ d^*_{\bf r} (\partial_\tau+i\lambda^{(1)}_{\bf
r}-i\sum_\sigma\lambda^{(2)}_{{\bf r}\sigma}+U) d_{\bf r} \Bigr ] .
\label{Sgam}
\end{eqnarray}
\eleq
where 
\begin{equation}
\gamma_{\bf r} ={ \gamma_{{\bf r}\uparrow} \choose  \gamma_{{\bf
r}\downarrow} } \equiv { \gamma_{{\bf r}-\uparrow} \choose \gamma_{{\bf
r}+\downarrow} }. 
\end{equation}
The fields $\gamma_{{\bf r}-\downarrow}$ and $\gamma_{{\bf
r}+\uparrow}$ do not appear any more. Note that the UHB (LHB) can be
labeled by the band index $\alpha=+$ ($\alpha=-$) or the spin index
$\sigma=\downarrow$ ($\sigma=\uparrow$). In the following we use the
spin index to label the Hubbard bands. $A^0_{{\bf r}z}= (R^\dagger_{\bf r}\dot
R_{\bf r})_{\uparrow\uparrow}$ is defined by writing the gauge field
$A^0_{\bf r}=R^\dagger_{\bf r}\dot
R_{\bf r}$ as $A^0_{\bf r}= \sum_{\nu=x,y,z} A^0_{{\bf r}\nu}\sigma_\nu$. 

The Hubbard-Stratonovich decoupling of the hopping term requires the
introduction of the auxiliary fermionic field
\begin{equation}
\psi_{\bf r} ={ \psi_{{\bf r}\uparrow} \choose  \psi_{{\bf
r}\downarrow} } \equiv { \psi_{{\bf r}-\uparrow} \choose \psi_{{\bf
r}+\downarrow} }. 
\end{equation}
The partition function can be written as
\begin{eqnarray}
Z &=& \int {\cal D}{\bf\Omega} \,{\rm det}(\hat t) \int {\cal D}[\psi]
\exp \Bigl \lbrace 
-\sum_{a,b} \psi^*_a \hat t^{-1}_{ab}\psi_b \Bigr \rbrace \nonumber \\
&& \times \int d\lambda \, \int {\cal D}[f,e,p,d] \exp \Bigl \lbrace
-S_{\rm at}+\sum_a (\psi^*_a \gamma_a + {\rm c.c.}) \Bigr \rbrace ,
\nonumber \\ && 
\end{eqnarray}
where the hopping matrix 
\begin{equation}
\hat t_{{\bf rr}'}=R^\dagger_{\bf r}t_{{\bf rr}'} R_{{\bf r}'}
\end{equation}
acts on the spin/band indices. Note that ${\rm det}(\hat t)$ depends on
$\bf\Omega$ and should therefore be kept explicitly in the functional
integral. Using
Eq.~(\ref{Rdef}), the matrix $R^\dagger_{\bf r} R_{{\bf r}'}$ can be
expressed as 
\bleq
\begin{equation}
R^\dagger_{\bf r} R_{{\bf r}'} = \frac{1}{\sqrt{2}}\left ( 
\begin{array}{cc} 
(1+{\bf\Omega}_{\bf r}\cdot {\bf\Omega}_{{\bf r}'})^{1/2} 
e^{i\Phi_{{\bf r},{\bf r}'}/2} & 
(1-{\bf\Omega}_{\bf r}\cdot {\bf\Omega}_{{\bf r}'})^{1/2}
e^{i\Phi'_{{\bf r},{\bf r}'}/2} \\ 
-(1-{\bf\Omega}_{\bf r}\cdot {\bf\Omega}_{{\bf r}'})^{1/2}
e^{-i\Phi'_{{\bf r},{\bf r}'}/2} & 
(1+{\bf\Omega}_{\bf r}\cdot {\bf\Omega}_{{\bf r}'})^{1/2}
e^{-i\Phi_{{\bf r},{\bf r}'}/2}
\end{array}
\right ) ,
\label{RR}
\label{RdR}
\end{equation}
\eleq
where $\Phi_{{\bf r},{\bf r}'}\equiv \Phi({\bf\Omega}_{\bf
r},{\bf\Omega}_{{\bf r}'})$ is the signed solid angle spanned by
the vectors ${\bf\Omega}_{\bf r},{\bf\Omega}_{{\bf r}'},\hat {\bf z}$,
and $\Phi'_{{\bf r},{\bf 
r}'}=\Phi({\bf\Omega}_{\bf r},-{\bf\Omega}_{{\bf r}'})-2\varphi_{{\bf
r}'}$. $\Phi$ and $\Phi'$ satisfy the relations $\Phi_{{\bf r}',{\bf
r}}= -\Phi_{{\bf r},{\bf r}'}$, $\Phi'_{{\bf r}',{\bf r}}= \Phi'_{{\bf
r},{\bf r}'}+2\pi$. 

Performing the integration over the fields $f,e,p,d$ and the Lagrange
multipliers, we obtain
\begin{eqnarray}
Z &=& \int {\cal D}{\bf\Omega}\, Z_{\rm at}[{\bf\Omega}]\,{\rm det}(\hat t) 
\nonumber \\ && \times \int {\cal D}[\psi]
 \exp \Bigl \lbrace -\sum_{a,b} \psi^*_a \hat t^{-1}_{ab}\psi_b
 +W[\psi^*,\psi;{\bf\Omega}] \Bigr \rbrace ,
\label{Spsi1}
\end{eqnarray}
where 
\begin{eqnarray}
Z_{\rm at}[{\bf\Omega}] &=& \int d\lambda \, \int {\cal D}[f,e,p,d]
\exp \Bigl \lbrace -S_{\rm at} \Bigr \rbrace \nonumber \\ &=& 
Z_{\rm at}^{(0)} \exp \Bigl \lbrace -S_B[{\bf\Omega}] \Bigr \rbrace
\label{SB1}
\end{eqnarray}
is the partition function in the atomic limit for a given
configuration of the fluctuating spin-quantization axis
${\bf\Omega}_{\bf r}$. The 
generating functional of the connected atomic Green's functions
$W[\psi^*,\psi;{\bf\Omega}]$ depends on ${\bf\Omega}_{\bf r}$ since
$S_{\rm at}$ does. $S_B[{\bf\Omega}]$ is calculated by treating the
gauge field $A^0_{{\bf r}z}$ in
perturbation in Eq.~(\ref{SB1}). To lowest order (in a cumulant expansion) 
\begin{eqnarray}
S_B[{\bf\Omega}] &=& \sum_{\bf r} \int d\tau\,
\langle f^*_{{\bf r}\uparrow} p^*_{{\bf r}\uparrow} f_{{\bf r}\uparrow}
p_{{\bf r}\uparrow} \rangle _{S_{\rm at}^{(0)}} A^0_{{\bf
r}z} \nonumber \\ &=& \sum_{\bf r} \int d\tau \,
A^0_{{\bf r}z} .
\label{SBi}
\end{eqnarray}
Higher-order corrections, of order $t/U$, are ignored. This is
justified since the spin dynamics is slow compared to the
interband-transition dynamics. [The second line of
Eq.~(\ref{SBi}) is easily obtained by following the ``philosophy'' of
Appendix \ref{appIII}.] $S_B$ is a collection of Berry phase terms for spins
localized at the lattice sites.\cite{Shapere} 
Using Eq.~(\ref{Rdef}), it can be written as
\begin{eqnarray}
S_B[{\bf\Omega}] &=& \frac{i}{2} \sum_{\bf r} \int d\tau\,
(1-\cos\theta_{\bf r})\dot \varphi_{\bf r} \nonumber \\
&=& \frac{i}{2} \sum_{\bf r} \int d\tau\, 
{\bf A}({\bf\Omega})\cdot \dot{\bf\Omega} ,
\label{SB0}
\end{eqnarray}
where ${\bf A}({\bf\Omega})$ is the vector potential created by a
magnetic monopole sitting at the center of a unit sphere.

Thus, the action of the $\psi$ field can be written as [see
Eqs.~(\ref{Spsi1}) and (\ref{SB1})]
\begin{eqnarray}
S[\psi^*,\psi;{\bf\Omega}] &=& -\ln {\rm det}(\hat t)+ S_B[{\bf\Omega}]
\nonumber \\ && 
+ \sum_{a,b} \psi^*_a \hat 
t^{-1}_{ab} \psi_b -W[\psi^*,\psi;{\bf\Omega}] \nonumber \\ &=&
-\ln {\rm det}(\hat t) + S_B[{\bf\Omega}] + 
\sum_{a,b} \psi^*_a (\hat t^{-1}_{ab}+G_{ab}) \psi_b \nonumber \\ && 
- \frac{1}{(2!)^2} \sum_{a_i,b_i} G^{{\rm
II}c}_{a_1a_2,b_1b_2} \psi^*_{a_1}\psi^*_{a_2}\psi_{b_2}\psi_{b_1} .
\label{Spsi2}
\end{eqnarray}
In the second line of (\ref{Spsi2}), we have retained only the
quadratic and quartic
parts, since this is sufficient to obtain the effective action
of holes in the LHB to first order in $t/U$ (see below). 

In order to  completely determine the action
$S[\psi^*,\psi;{\bf\Omega}]$, we need to compute $G$ and $G^{{\rm
II}c}$. The single-particle Green's function $G$ is calculated in
Appendix \ref{appIII}: 
\begin{equation}
G^{-1}_{{\bf r}\sigma} = G^{(0)-1}_\sigma-{\rm sgn}(\sigma) 
A^0_{{\bf r}z} ,  
\label{Gat1}
\end{equation}
where ${\rm sgn}(\sigma)=1(-1)$ for $\sigma=\uparrow(\downarrow)$. 
$G^{(0)}$ is the atomic Green's function corresponding to
$S_{\rm at}^{(0)}$ [Eq.~(\ref{Sgam})]:
\begin{eqnarray}
G^{(0)}_\uparrow(\tau) &=&  \theta(-\tau+\eta) e^{\mu\tau} , \nonumber \\ 
G^{(0)}_\downarrow(\tau) &=& -\theta(\tau-\eta) e^{(\mu-U)\tau} ,
\label{G0} 
\end{eqnarray}
where $\eta\to 0^+$, and the limit $T\to 0$ has been taken. 
In Fourier space, Eqs.~(\ref{G0}) become $G^{(0)}_\uparrow(i\omega)
=(i\omega+\mu)^{-1}$ and $G^{(0)}_\downarrow(i\omega)=(i\omega+\mu-U)^{-1}$.

Neglecting the Berry phase term $A^0$, we can obtain $G^{{\rm II}c}$
using the method of Appendix \ref{appI}. It is straightforward to
modify this approach in order to impose the constraint that the LHB is
populated by up-spin particles only. At $T=0$, we find
\begin{eqnarray}
G^{{\rm
II}c}_{\sigma\sigma,\sigma\sigma}(\tau_1,\tau_2;\tau_3,\tau_4)&=&0, 
\label{GIIssss} \\ 
G^{{\rm II}c}_{\uparrow\downarrow,\uparrow\downarrow}
(\tau_1,\tau_2;\tau_3,\tau_4) &=& G_\uparrow^{(0)}(\tau_1-\tau_3)
G_\downarrow^{(0)}(\tau_2-\tau_4) \nonumber \\ && \times 
[\theta(\tau_1-\tau_2)-\theta(\tau_3-\tau_4)] .
\end{eqnarray}
Since $G^{{\rm II}c}_{\sigma\sigma,\sigma\sigma}=0$, $G^{{\rm II}c}$
necessarily involves (virtual) interband transitions. The typical
energy scale ($\sim U$) for these transitions being much larger than
the typical energy scale for spin fluctuations ($\sim J=4t^2/U$), we
can ignore the effect of $A^0$ on $G^{{\rm II}c}$. Eq.~(\ref{GIIssss})
can easily be extended to higher-order connected Green's functions:
\begin{equation}
G^{Rc}_{\sigma\cdots\sigma,\sigma\cdots\sigma}=0 \,\,\,\,\,\, (R\geq 2). 
\end{equation}
In the limit $U\to\infty$, where
interband transitions are suppressed, the action of the $\psi$ field is
Gaussian. It remains nevertheless highly non-trivial since the
intersite hopping term
depends on the spin variables ${\bf\Omega}_{\bf r}$. [This limit is
discussed in Sec.~\ref{sec:Uinf}.] Furthermore, this implies that only
a limited number of higher-order vertices $G^{Rc}$ ($R\geq 2$) is
necessary to obtain the effective action to a given order in $t/U$ (if
one considers only the quadratic and quartic parts of the effective action),
since these vertices necessarily involve interband transitions. In
particular, it is sufficient to consider $G^{{\rm II}c}$ to obtain the
effective action to first order in $t/U$. This is an important
difference with the approach of Sec.~\ref{sec:sce}, where the action of
the auxiliary fermionic field $\psi$ involves an infinite number of
vertices even in the limit $U\to\infty$. 

From Eq.~(\ref{Spsi2}), it is now possible to integrate out the UHB to
obtain the effective action of a hole in the LHB. It is more
convenient to first perform another Hubbard-Stratonovich
transformation of the hopping term in (\ref{Spsi2}). If this
transformation were carried out exactly (i.e. on the exact action
$S[\psi^*,\psi;{\bf\Omega}]$), one would essentially ``undo'' the first
Hubbard-Stratonovich transformation and  recover the original action
$S[\gamma^*,\gamma;{\bf\Omega}]$ [Eq.~(\ref{Sgam})].  Here the idea is
that to first order in $t/U$, one can truncate the action
$S[\psi^*,\psi;{\bf\Omega}]$ as in Eq.~(\ref{Spsi2}). One then obtains
an effective action $S[\gamma^*,\gamma;{\bf\Omega}]$ which contains
all processes of order $t/U$. It is then possible to integrate out the
UHB (i.e. the field $\gamma_\downarrow$) to obtain the effective action
$S[\gamma^*_\uparrow,\gamma_\uparrow;{\bf\Omega}]$ for the LHB. 

Following this procedure, we obtain
\begin{eqnarray}
Z &=& \int {\cal D}{\bf\Omega} \int {\cal D}[\gamma] \, \exp \Bigl
\lbrace -S[\gamma^*,\gamma;{\bf\Omega}] \Bigr \rbrace , \nonumber \\ 
S[\gamma^*,\gamma;{\bf\Omega}] &=& S_B[{\bf\Omega}] - \sum_{a,b}
\gamma^*_a \hat t_{ab} \gamma_b \nonumber \\ &&
- \tilde W[\gamma^*,\gamma;{\bf\Omega}] - \ln
\tilde Z[{\bf\Omega}] , 
\end{eqnarray}
where 
\begin{equation}
\tilde Z[{\bf\Omega}]  = \int {\cal D}[\psi] \exp \Bigl \lbrace
W[\psi^*,\psi;{\bf\Omega}] \Bigr \rbrace .
\end{equation}
Note that $\gamma$, which is the auxiliary Grassmannian field of the
Hubbard-Stratonovich transformation, is now an independent variable. 
$\tilde W[\gamma^*,\gamma;{\bf\Omega}]$ is the generating functional
of connected Green's functions obtained from the action
$-W[\psi^*,\psi;{\bf\Omega}]$.
To lowest order in $t/U$, $\ln \tilde Z[{\bf\Omega}]=\ln {\rm det}(G)
= S_B[{\bf\Omega}]$. Again we neglect corrections of order $t/U$ to
the Berry phase term. Retaining only the quadratic and quartic parts
of the effective action, we have
\begin{eqnarray}
S[\gamma^*,\gamma;{\bf\Omega}] &=&  - \sum_{a,b} \gamma^*_a (\hat
t_{ab} -\tilde G_{ab}) \gamma_b \nonumber \\ && 
- \frac{1}{(2!)^2} \sum_{a_i,b_i} \tilde G^{{\rm
II}c}_{a_1a_2,b_1b_2} 
\gamma^*_{a_1} \gamma^*_{a_2} \gamma_{b_2} \gamma_{b_1} ,
\label{Gtilde0}
\end{eqnarray}
where
\begin{eqnarray}
\tilde G_{ab} &=& - \langle \psi_a\psi^*_b \rangle _{-W} , \nonumber \\ 
\tilde G^{{\rm II}c}_{a_1a_2,b_1b_2} &=& \langle
\psi_{a_1}\psi_{a_2}\psi^*_{b_2}\psi^*_{b_1} \rangle _{-W,c} .
\end{eqnarray}
In Eq.~(\ref{Gtilde0}), it is sufficient to determine $\tilde G$ and
$\tilde G^{{\rm II}c}$ to first order in $t/U$. We find
\begin{eqnarray}
\tilde G_{ab} &=& -G^{-1}_{ab} + \tilde G^{(1)}_{ab} , \nonumber \\ 
\tilde G^{{\rm II}c}_{a_1a_2,b_1b_2} &=& \sum_{a'_i,b'_i}
G^{-1}_{a_1a'_1} G^{-1}_{a_2a'_2} G^{{\rm II}c}_{a'_1a'_2,b'_1b'_2}
G^{-1}_{b'_1b_1} G^{-1}_{b'_2b_2} , 
\label{Gtilde1}
\end{eqnarray}
where 
\begin{equation}
\tilde G^{(1)}_{ab} = \sum_{a'_i,b'_i} G^{-1}_{aa'_1} 
G^{{\rm II}c}_{a'_1a'_2,b'_1b'_2} G^{-1}_{b'_2a'_2}G^{-1}_{b'_1b}  
\end{equation}
is the ``one-loop'' correction to $-G^{-1}$. From (\ref{Gtilde1}), we
conclude that $\tilde G^{{\rm II}c}$ is simply related to the
atomic two-particle vertex: 
\begin{equation}
\tilde G^{{\rm II}c}=-\Gamma^{\rm II} .
\end{equation}
We thus  obtain
\begin{eqnarray}
S[\gamma^*,\gamma;{\bf\Omega}] &=&  - \sum_{a,b} \gamma^*_a (\hat
t_{ab} + G^{-1}_{ab}-\tilde G^{(1)}_{ab}) \gamma_b \nonumber \\ && 
+ \frac{1}{(2!)^2} \sum_{a_i,b_i} \Gamma^{\rm II}_{a_1a_2,b_1b_2}
\gamma^*_{a_1} \gamma^*_{a_2} \gamma_{b_2} \gamma_{b_1} .
\label{Sgam0}
\end{eqnarray}
Note the important role played by $\tilde G^{(1)}$. If we consider the
one-loop correction due to $\Gamma^{\rm II}$, we obtain (among other
contributions) an atomic contribution to the propagator of the field
$\gamma$, which is already included in $G$. This contribution is
precisely canceled by $\tilde G^{(1)}$ (see section \ref{subsubsec:t/U}). 

Eq.~(\ref{Sgam0}) has a clear physical meaning. By performing two
successive Hubbard-Stratonovich transformations, and truncating the
action $S[\psi^*,\psi;{\bf\Omega}]$ in the intermediate step, we have
effectively summed the atomic contributions to the single-particle
propagator and the two-particle vertex. Since $\Gamma^{\rm
II}_{\sigma\sigma,\sigma\sigma}=0$, perturbative corrections (due to
$\Gamma^{\rm II}_{\uparrow\downarrow,\uparrow\downarrow}$) to the
Gaussian action are of order $t/U$. In
the limit $U\to\infty$, $S[\gamma^*,\gamma;{\bf\Omega}]$ reduces to
its Gaussian part (except the term $\gamma^*\tilde G^{(1)}\gamma$ which is
of order $t/U$). The effective action for a hole in the LHB can be
derived by integrating out the UHB, the two-particle vertex
$\Gamma^{\rm II}_{\uparrow\downarrow,\uparrow\downarrow}$ being
considered within perturbation theory.

\subsubsection{Limit $U\to\infty$} 
\label{sec:Uinf}

In the absence of interband coupling, we obtain the
following effective action for the LHB 
\begin{eqnarray}
&& S^{(0)}_{\rm LHB} = \sum_{\bf r} \int d\tau \, \gamma^*_{\bf r} 
(\partial_\tau -\mu+A^0_{{\bf r}z})\gamma_{\bf r}
\nonumber \\ && 
- \frac{t}{\sqrt{2}} \sum_{\langle{\bf r},{\bf r}'\rangle} \int d\tau
\, (1+{\bf\Omega}_{\bf r}\cdot {\bf\Omega}_{{\bf r}'})^{1/2}
[\gamma^*_{\bf r} \gamma_{{\bf r}'} 
e^{i\Phi_{{\bf r},{\bf r}'}/2} + {\rm c.c.} ] , \nonumber \\ &&
\label{Sgam1}
\end{eqnarray}
where we have used Eqs.~(\ref{RdR}), (\ref{Gat1}) and (\ref{G0}). 
From now on, we drop the spin index for fermions in the LHB and denote
by $\gamma$ the field $\gamma_\uparrow$. 
The action (\ref{Sgam1}) is similar to that obtained by Schulz from the
large-$U$ Hartree-Fock saddle point. \cite{Schulz90} 
At half-filling ($\gamma^*_{{\bf r}\uparrow}\gamma_{{\bf
r}\uparrow}=1$), it reduces to $S_B[{\bf\Omega}]$. As expected,
the half-filled Hubbard model becomes a collection of independent
spins in the absence of interband coupling ($U\to\infty$).
The coupling between a hole in the LHB and spin fluctuations is
described by a U(1) gauge-field theory. 
\cite{Baskaran88,Wiegmann88,Wen89,Lee89,Shankar89,Schulz90} As pointed out in
Ref.~\onlinecite{Schulz90}, the introduction of a hole (absence of a
$\gamma_\uparrow$ particle) in the LHB has two effects. (i) The Berry
phase term $A^0_{{\bf r}z}\gamma^*_{\bf r}\gamma_{\bf r}$, 
which is alive when the site ${\bf r}$ is occupied by a
particle, is suppressed by the introduction of a hole. This
conclusion was first obtained by Shankar from semi-phenomenological
arguments. \cite{Shankar89}  (ii) The 
hole can propagate through the system, but spin fluctuations
modify the kinetic energy term. The hopping term between two
neighboring sites $\bf r$ and ${\bf r}'$ has its amplitude
reduced by a factor $[(1+{\bf\Omega}_{\bf r}\cdot{\bf\Omega}_{{\bf
r}'})/2]^{1/2}$, and acquires the phase $\Phi_{{\bf r},{\bf r}'}/2$. 
The latter can be
interpreted as resulting from an effective magnetic field depending on
the spin configuration. \cite{Doucot89} It is then equal to the
circulation of the magnetic vector potential on the link between $\bf r$ and
${\bf r}'$. When going around an elementary plaquette, the accumulated
phase $\Phi_{1234}$ corresponds to the magnetic flux through the
plaquette, and is given by half the solid
angle spanned by the unit vectors
${\bf\Omega}_1,{\bf\Omega}_2,{\bf\Omega}_3,{\bf\Omega}_4$:
\begin{equation}
\Phi_{1234}=\frac{1}{2}\bigl [ \Phi_{1,2}+\Phi_{2,3}
+\Phi_{3,4}+\Phi_{4,1} \bigr ] . 
\end{equation}
We expect the kinetic energy term to be optimized for a ferromagnetic
configuration of the spins, since in that case the effective magnetic
field vanishes and the hopping amplitude takes on its maximum
value. This is nothing but the familiar Nagaoka
phenomenon.\cite{Nagaoka66,Schulz90,Doucot89}

\subsubsection{Correction of order $t/U$} 
\label{subsubsec:t/U}

The integration of the field $\gamma_\downarrow$ in Eq.~(\ref{Sgam0})
will generate a correction $S^{(1)}_{\rm LHB}$ to the effective action 
$S^{(0)}_{\rm LHB}$. Consider first the quadratic part of
$S^{(1)}_{\rm LHB}$. To first order in $t/U$, it is given by
$S^{(1a)}_{\rm LHB}+S^{(1b)}_{\rm LHB}$, where
\bleq
\begin{eqnarray}
S^{(1a)}_{\rm LHB} &=& \sum_{{\bf r},{\bf r}',{\bf r}''} \int d\tau
d\tau'\, \gamma^*_{\bf r}(\tau) (\hat t_{{\bf
rr}'}(\tau))_{\uparrow\downarrow} G_\downarrow(\tau,\tau') 
(\hat t_{{\bf r}'{\bf r}''}(\tau'))_{\downarrow\uparrow}
\gamma_{{\bf r}''}(\tau') ,  \label{S1a} \\
S^{(1b)}_{\rm LHB} &=& \sum_{\bf r} \int
d\tau d\tau'd\tau_2d\tau_4\, \gamma^*_{\bf r}(\tau) 
\gamma_{\bf r}(\tau') \Gamma^{\rm
II}_{\uparrow\downarrow,\uparrow\downarrow} (\tau,\tau_2;\tau',\tau_4)
(G^{-1}_\downarrow+\hat t_{\downarrow\downarrow})^{-1}_{{\bf
r}\tau_4,{\bf r}\tau_2} .
\end{eqnarray}
\eleq
$S^{(1b)}_{\rm LHB}$ can be expanded in a series in powers of $\hat
t_{\downarrow\downarrow}$. The lowest-order term exactly cancels the
term $\gamma^*_\uparrow\tilde G^{(1)}_\uparrow\gamma_\uparrow$ in
Eq.~(\ref{Sgam0}). The next-order term ($O(\hat
t^2_{\downarrow\downarrow}))$ is proportional to $\int
d\tau_2d\tau_4\, G^{{\rm II}c}_{\uparrow\downarrow,\uparrow\downarrow}
(\tau_1,\tau_2;\tau_3,\tau_4)G_\downarrow(\tau_4,\tau_2)$  and
therefore vanishes [since $G^{{\rm II}c}_{\uparrow\downarrow, 
\uparrow\downarrow}(\tau_1,\tau_2;\tau_3,\tau_4) \propto
G_\downarrow(\tau_2,\tau_4)$]. For a similar reason, all higher-order terms
vanish. Thus the quadratic part of the effective action reduces to
$S^{(0)}_{\rm LHB}+S^{(1a)}_{\rm LHB}$.

\begin{figure}
\epsfxsize 8.cm 
\epsffile[45 375 300 600]{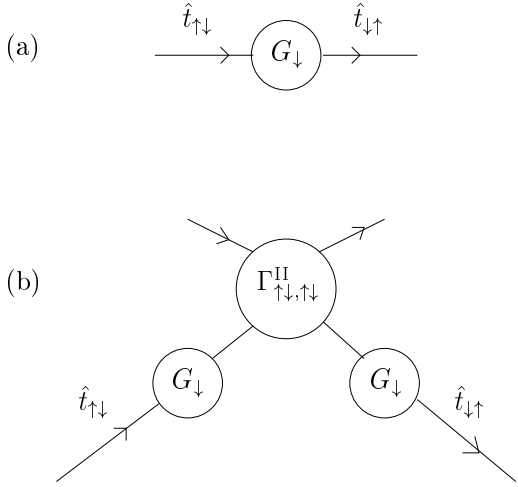}
\caption{ 
Diagrammatic representation of the contribution of order $t/U$
to the effective action $S_{\rm LHB}$. (a) Quadratic part
$S^{(1a)}_{\rm LHB}$. (b) Quartic part $S^{(1c)}_{\rm LHB}$. When
written in a time-ordered fashion, $S^{(1c)}_{\rm LHB}$ generates a
quadratic term which exactly cancels $S^{(1a)}_{\rm LHB}$ (see text). }
\label{Fig1}
\end{figure}

Consider now $S^{(1a)}_{\rm LHB}$ (shown diagrammatically in
Fig.~\ref{Fig1}(a)). We have argued above that we can
ignore the effect of spin fluctuations on (virtual) interband
transitions. This allows us to replace $G_\downarrow(\tau,\tau')$ by 
$G^{(0)}_\downarrow(\tau-\tau')$ and $(\hat t_{{\bf r}'{\bf
r}''}(\tau'))_{\uparrow\downarrow}$ by $(\hat t_{{\bf r}'{\bf
r}''}(\tau))_{\uparrow\downarrow}$ in Eq.~(\ref{S1a}). Furthermore, since the
characteristic energy scale ($\sim U$) of a virtual transition is
larger than any other typical energy in the system, we can assume
$S^{(1a)}_{\rm LHB}$ to be local in time, i.e. $\gamma^*_{\bf
r}(\tau)\gamma_{{\bf r}'}(\tau')\simeq  \gamma^*_{\bf
r}(\tau)\gamma_{{\bf r}'}(\tau)$. The simple replacement
$\gamma_{{\bf r}'}(\tau')\to \gamma_{{\bf r}'}(\tau)$
would however lead to a wrong result. Instead, we write
$\gamma(\tau')=\gamma(\tau)e^{-\mu(\tau-\tau')}$, which follows from
the equation of motion for the field $\gamma$ in the atomic
limit. \cite{Zinn-Justin} $S^{(1a)}_{\rm LHB}$ is then approximated as
\begin{eqnarray}
S^{(1a)}_{\rm LHB} &=& \sum_{{\bf r},{\bf r}',{\bf r}''} \int d\tau \,
\gamma^*_{\bf r}(\tau) \gamma_{{\bf r}''}(\tau)
(\hat t_{{\bf rr}'}(\tau))_{\uparrow\downarrow}
(\hat t_{{\bf r}'{\bf r}''}(\tau))_{\downarrow\uparrow}
\nonumber \\ && \times 
\int d\tau'\,  G^{(0)}_\downarrow(\tau-\tau') e^{-\mu(\tau-\tau')} .
\end{eqnarray}
Using Eq.~(\ref{G0}),  the sum over $\tau'$ can be written as
\begin{equation}
- \int_{0}^\tau d\tau' \, e^{-U(\tau-\tau')} \simeq - \frac{1}{U} .
\end{equation}
Using Eq.~(\ref{RR}), we finally deduce
\bleq 
\begin{eqnarray}
S^{(1a)}_{\rm LHB} &=&
-\frac{t^2}{2U} \sum_{\langle {\bf r},{\bf r}' \rangle} \int d\tau\,
[\gamma^*_{\bf r}\gamma_{\bf r}
+\gamma^*_{{\bf r}'}\gamma_{{\bf r}'} ] 
(1-{\bf\Omega}_{\bf r}\cdot {\bf\Omega}_{{\bf r}'})
\nonumber \\ && 
-\frac{t^2}{2U} \sum_{\langle {\bf r},{\bf r}',{\bf r}''\rangle} \int
d\tau \, 
(1-{\bf\Omega}_{\bf r}\cdot{\bf\Omega}_{{\bf r}'})^{1/2} 
(1-{\bf\Omega}_{{\bf r}'}\cdot{\bf\Omega}_{{\bf r}''})^{1/2} 
[\gamma^*_{\bf r}\gamma_{{\bf r}''}
e^{i(\Phi'_{{\bf r},{\bf r}'}-\Phi'_{{\bf r}'',{\bf r}'})/2} 
+ {\rm c.c.} ] , 
\end{eqnarray}
\eleq 
Here $\langle {\bf r},{\bf r}',{\bf r}''\rangle$ stands for a
three-site term with both $\langle {\bf r},{\bf r}'\rangle$ and
$\langle {\bf r}',{\bf r}''\rangle$ first neighbors (${\bf r}\neq {\bf
r}''$). 

The effective action $S^{(0)}_{\rm LHB}+S^{(1a)}_{\rm LHB}$ is similar
to that of Ref.~\onlinecite{Schulz90} derived by studying
fluctuations around the large-$U$ Hartree-Fock saddle point. In the
absence of  holes in the LHB, it reduces to the action the
AF Heisenberg model with coupling constant $J=4t^2/U$,
\begin{equation}
S_{\rm Heis}[{\bf\Omega}] =
S_B[{\bf\Omega}]+J\sum_{\langle {\bf r},{\bf r}' \rangle}
\int d\tau\, \Bigl(\frac{{\bf\Omega}_{\bf r}\cdot
{\bf\Omega}_{{\bf r}'}}{4}-\frac{1}{4} \Bigr) ,
\label{Heis}
\end{equation}
which is the expected result. Nevertheless, we will show that
$S^{(0)}_{\rm LHB}+S^{(1a)}_{\rm LHB}$ is {\it not} the effective
action to first order in $t/U$. One has to consider the quartic
contribution $S^{(1c)}_{\rm LHB}$ to $S^{(1)}_{\rm LHB}$ which results
from the integration of the UHB. It will turn out that
$S^{(1c)}_{\rm LHB}$, when written in a time-ordered fashion (as it should be
in the functional integral), generates a quadratic term which exactly
cancels $S^{(1a)}_{\rm LHB}$. What will be left is nothing but the
action of the $t$-$J$ model. In fact, it is easily seen on physical
grounds that 
$S^{(1a)}_{\rm LHB}$ is not the action that we expect to first order in
$t/U$. The introduction of a hole at a given site $\bf r$ should suppress the
AF exchange with all the neighboring sites. Now, according to
$S^{(1a)}_{\rm LHB}$, the AF coupling between two neighboring sites
$\bf r$ and ${\bf r}'$ is reduced by the presence of a hole at site
$\bf r$ or ${\bf r}'$, but vanishes only if two holes, one at each
site, are introduced in the system. 

The quartic contribution to the action, shown in Fig.~\ref{Fig1}(b), is
given by
\bleq
\begin{eqnarray}
S^{(1c)}_{\rm LHB} &=& \sum_{{\bf r},{\bf r}',{\bf r}''} \int
d\tau_1d\tau_2d\tau_3d\tau_4 \int d\tau_2'd\tau_4'\, \Gamma^{\rm
II}_{\uparrow\downarrow,\uparrow\downarrow}(\tau_1,\tau_2';\tau_3,\tau_4') 
G^{(0)}_\downarrow(\tau_2-\tau_2')G^{(0)}_\downarrow(\tau_4'-\tau_4)
\nonumber \\ && \times 
(\hat t_{{\bf r}'{\bf r}}(\tau_2))_{\uparrow\downarrow} 
(\hat t_{{\bf rr}''}(\tau_4))_{\downarrow\uparrow} 
\gamma^*_{\bf r}(\tau_1) \gamma^*_{{\bf r}'}(\tau_2) 
\gamma_{{\bf r}''}(\tau_4) \gamma_{\bf r}(\tau_3) ,
\label{S1c1}
\end{eqnarray}
where 
\begin{equation}
\Gamma^{\rm II}
_{\uparrow\downarrow,\uparrow\downarrow}(\tau_1,\tau_2;\tau_3,\tau_4)  
 = G^{(0)-1}_\uparrow(\tau_1-\tau_3)G^{(0)-1}_\downarrow(\tau_2-\tau_4)
[\theta(\tau_1-\tau_2)-\theta(\tau_3-\tau_4)].
\end{equation}
As for $S^{(1a)}_{\rm LHB}$, one can assume $S^{(1c)}_{\rm LHB}$ to be
local in time. Using $\gamma(\tau)\simeq \gamma(0)e^{\mu\tau}$ and 
$\gamma^*(\tau)\simeq \gamma^*(0)e^{-\mu\tau}$, we obtain
\begin{eqnarray}
S^{(1c)}_{\rm LHB} &=& \sum_{{\bf r},{\bf r}',{\bf r}''} \int d\tau_1\, 
\gamma^*_{\bf r}(\tau_1) \gamma^*_{{\bf r}'}(\tau_1) 
\gamma_{{\bf r}''}(\tau_1) \gamma_{\bf r}(\tau_1+\eta)
(\hat t_{{\bf r}'{\bf r}}(\tau_1))_{\uparrow\downarrow} 
(\hat t_{{\bf rr}''}(\tau_1))_{\downarrow\uparrow} \nonumber \\ && \times
\int d\tau_2d\tau_3d\tau_4\, e^{-\mu(\tau_1+\tau_2-\tau_3-\tau_4)} 
G^{(0)-1}_\uparrow(\tau_1-\tau_3)G^{(0)}_\downarrow(\tau_2-\tau_4) 
[\theta(\tau_1-\tau_4)-\theta(\tau_3-\tau_2)] ,
\label{S1c2}
\end{eqnarray}
where $\eta\to 0^+$. This equation has been obtained by noting that
\begin{equation}
G^{(0)-1}_\uparrow(\tau) = \mu\delta(\tau+\eta)-\dot\delta(\tau+\eta) ,
\label{G0-1}
\end{equation}
where the overdot denotes time derivative. 
The infinitesimal $\eta$, which is required by a proper treatment of
equal-time correlation functions, is crucial here. It implies that
$\tau_3$ is larger than $\tau_1$ in Eq.~(\ref{S1c1}). When
approximating $S^{(1c)}_{\rm LHB}$ by a local vertex (in time space),
one should keep track of this time ordering, since it does not
correspond to the one implicitly assumed in the functional integral.
Performing the sum over $\tau_2,\tau_3,\tau_4$ in Eq.~(\ref{S1c2}), we deduce
\begin{equation}
S^{(1c)}_{\rm LHB} = - \frac{1}{U} \sum_{{\bf r},{\bf r'},{\bf r}''}
\int d\tau \, 
\gamma^*_{\bf r}(\tau)
\gamma_{\bf r}(\tau+\eta)
\gamma^*_{{\bf r}'}(\tau) 
\gamma_{{\bf r}''}(\tau) 
(\hat t_{{\bf r}'{\bf r}}(\tau))_{\uparrow\downarrow} 
(\hat t_{{\bf rr}''}(\tau))_{\downarrow\uparrow} .
\end{equation}
Noting that $\gamma^*_{\bf r}(\tau) 
\gamma_{\bf r}(\tau+\eta)\equiv\gamma^*_{\bf r}(\tau) 
\gamma_{\bf r}(\tau)-1$, we eventually come to
\begin{equation}
S^{(1c)}_{\rm LHB} = -S^{(1a)}_{\rm LHB}
- \frac{1}{U} \sum_{{\bf r},{\bf r'},{\bf r}''}
\int d\tau \, 
\gamma^*_{\bf r}
\gamma_{\bf r}
\gamma^*_{{\bf r}'} 
\gamma_{{\bf r}''} 
(\hat t_{{\bf r}'{\bf r}})_{\uparrow\downarrow} 
(\hat t_{{\bf rr}''})_{\downarrow\uparrow} .
\end{equation}
The quadratic contribution which is generated by the correct
time-ordering in the functional integral cancels $S^{(1a)}_{\rm
LHB}$. 

The effective action to first order in $t/U$, $S^{(0)}_{\rm
LHB}+S^{(1a)}_{\rm LHB}+S^{(1c)}_{\rm LHB}$, is therefore given by
\begin{eqnarray}
S_{\rm LHB} &=& \sum_{\bf r} \int d\tau \, \gamma^*_{\bf r} 
(\partial_\tau -\mu+A^0_{{\bf r}z})\gamma_{\bf r}
- \frac{t}{\sqrt{2}} \sum_{\langle{\bf r},{\bf r}'\rangle} \int d\tau
\, (1+{\bf\Omega}_{\bf r}\cdot {\bf\Omega}_{{\bf r}'})^{1/2}
[\gamma^*_{\bf r} \gamma_{{\bf r}'} 
e^{i\Phi_{{\bf r},{\bf r}'}/2} + {\rm c.c.} ] \nonumber \\ &&
+ J \sum_{\langle{\bf r},{\bf r}'\rangle} \int d\tau\, \Bigl
( \frac{{\bf\Omega}_{\bf r}\cdot{\bf\Omega}_{{\bf r}'}}{4}-\frac{1}{4}
\Bigr ) \gamma^*_{\bf r} 
\gamma_{\bf r}
\gamma^*_{{\bf r}'} 
\gamma_{{\bf r}'} \nonumber \\ && 
- \frac{t^2}{2U} \sum_{\langle{\bf r}',{\bf r},{\bf r}''\rangle} \int
d\tau \, (1-{\bf\Omega}_{\bf r}\cdot {\bf\Omega}_{{\bf r}'})^{1/2}
 (1-{\bf\Omega}_{\bf r}\cdot {\bf\Omega}_{{\bf r}''})^{1/2}
[e^{i(\Phi'_{{\bf r},{\bf r}'}-\Phi'_{{\bf r},{\bf r}''})/2} 
\gamma^*_{\bf r} \gamma_{\bf r}
\gamma^*_{{\bf r}'} 
\gamma_{{\bf r}''}  + {\rm c.c.} ] .
\label{Sfinal}
\end{eqnarray} 
The action (\ref{Sfinal}) exhibits the correct physical properties. 
The Berry phase term and the modification of the kinetic energy term
by spin fluctuations have been discussed in Sec.~\ref{sec:Uinf}. The
$J$-term describes AF exchange interactions
between neighboring sites $\langle {\bf r},{\bf r}'\rangle$. It
vanishes as soon as a hole is present at 
site $\bf r$ or ${\bf r}'$. The three-site term is also familiar from
the derivation of the $t$-$J$ model.

\subsubsection{Relation with other formalisms}     

In this section, we show that the action (\ref{Sfinal}) corresponds to
the action of the $t$-$J$ model obtained in the spin-hole coherent-state
path integral.\cite{Auerbach91,Auerbach} We then make the connection
with the slave-fermion formalism. 

We first perform a particle-hole transformation: $\gamma_{\bf r}\to
h^*_{\bf r}$, $\gamma^*_{\bf r}\to h_{\bf r}$. The action
(\ref{Sfinal}) then becomes
\begin{eqnarray}
S_{\rm LHB} &=& \sum_{\bf r} \int d\tau \, [h^*_{\bf r}(\partial_\tau
-\mu_h) h_{\bf r}+ A^0_{{\bf r}z}(1-h^*_{\bf r}h_{\bf r}) ] 
\nonumber \\ &&  
+ \frac{t}{\sqrt{2}} \sum_{\langle{\bf r},{\bf r}'\rangle} \int
d\tau\, (1+{\bf\Omega}_{\bf r}\cdot {\bf\Omega}_{{\bf r}'})^{1/2} 
[h^*_{\bf r}h_{{\bf r}'}e^{-i\Phi_{{\bf r},{\bf r}'}/2} + {\rm c.c.}] 
\nonumber \\  &&
+ J \sum_{\langle{\bf r},{\bf r}'\rangle} \int d\tau\, \Bigl
( \frac{{\bf\Omega}_{\bf r}\cdot{\bf\Omega}_{{\bf r}'}}{4}-\frac{1}{4}
\Bigr ) (1-h^*_{\bf r} h_{\bf r})
(1-h^*_{{\bf r}'} h_{{\bf r}'}) \nonumber \\ && 
+ \frac{t^2}{2U} \sum_{\langle{\bf r}',{\bf r},{\bf r}''\rangle} \int
d\tau \, (1-{\bf\Omega}_{\bf r}\cdot {\bf\Omega}_{{\bf r}'})^{1/2}
 (1-{\bf\Omega}_{\bf r}\cdot {\bf\Omega}_{{\bf r}''})^{1/2}
\nonumber \\ && \times 
[e^{i(\Phi'_{{\bf r},{\bf r}'}-\Phi'_{{\bf r},{\bf r}''})/2} 
(1-h^*_{\bf r} h_{\bf r})
h^*_{{\bf r}''} h_{{\bf r}'}  + {\rm c.c.} ] .
\label{Sfinal1}
\end{eqnarray} 
\eleq
where $\mu_h=-\mu$ is the chemical potential of the holes.
The action (\ref{Sfinal1}) is precisely the action of the $t$-$J$ model
in the spin-hole coherent-state path integral.\cite{Auerbach91,Auerbach}
Note that the three-site term (the so-called ``pair-hopping'' term, last
term of the rhs of Eq.~(\ref{Sfinal1})) is usually omitted in the
$t$-$J$ model.  

The rotation matrices $R_{\bf r}$ are elements of SU(2)/U(1). Instead of
integrating over this manifold in the functional integral, one can
choose to integrate over all SU(2) matrices (with the properly
normalized integration measure),\cite{Schulz90} i.e. one writes
\begin{equation}
R_{\bf r} = \left ( 
\begin{array}{cc} 
z_{1{\bf r}} & -z^*_{2{\bf r}} \\
 z_{2{\bf r}} & z^*_{1{\bf r}}
\end{array}
\right ),
\end{equation}
with the constraints 
\begin{equation}
|z_{1{\bf r}}^2|+|z_{2{\bf r}}^2|=1 .
\label{cons}
\end{equation} 
Writing
$z_{\bf r}=(z_{1{\bf r}},z_{2{\bf r}})^T$, one then has
\begin{eqnarray}
&& {\bf\Omega}_{\bf r} = z^\dagger_{\bf r}\bbox{\sigma}z_{\bf r} ,
\nonumber \\ &&
A^0_{{\bf r}z} = \frac{1}{2}(z^\dagger_{\bf r}\dot z_{\bf r}- \dot
z^\dagger_{\bf r}z_{\bf r}) , \nonumber \\  &&
\frac{1}{\sqrt{2}}(1+{\bf\Omega}_{\bf r}\cdot {\bf\Omega}_{{\bf
r}'})^{1/2} e^{i\Phi_{{\bf r},{\bf r}'}/2} 
= z^\dagger_{\bf r} z_{{\bf r}'} . 
\end{eqnarray}
Note that we have written $A^0_z$ in a symmetric way. Introducing new
variables (Schwinger bosons) defined by
\begin{equation}
b_{\bf r} = z_{\bf r}\Bigl ( 1-\frac{1}{2}h^*_{\bf r}h_{\bf r}\Bigr ),
\end{equation}
the constraints (\ref{cons}) become 
\begin{equation}
|b_{1{\bf r}}^2|+|b_{2{\bf r}}^2|+h^*_{\bf r}h_{\bf r} = 1 ,
\label{cons1}
\end{equation}
and the action reads
\bleq
\begin{eqnarray}
S_{\rm LHB} &=& \sum_{\bf r} \int d\tau \, [h^*_{\bf r}(\partial_\tau
-\mu_h) h_{\bf r}+ \frac{1}{2}(b^\dagger_{\bf r}\dot b_{\bf r}- \dot
b^\dagger_{\bf r}b_{\bf r} )]
+ t \sum_{\langle{\bf r},{\bf r}'\rangle} \int d\tau\, [
b^\dagger_{{\bf r}'} b_{\bf r}
h^*_{\bf r}h_{{\bf r}'} + {\rm c.c.}] 
\nonumber \\  &&
+ \frac{J}{4} \sum_{\langle{\bf r},{\bf r}'\rangle} \int d\tau\, 
[ b^\dagger_{\bf r} \bbox{\sigma} b_{\bf r}\cdot b^\dagger_{{\bf r}'}
\bbox{\sigma} b_{{\bf r}'} - (1-h^*_{\bf r}h_{\bf r}) (1-h^*_{{\bf
r}'}h_{{\bf r}'}) ] ,
\label{Sfinal2}
\end{eqnarray} 
\eleq 
where, for simplicity, we have omitted the ``pair-hopping'' term. The action
(\ref{Sfinal2}), together with the constraints (\ref{cons1}), is the
action of the $t$-$J$ model in the slave-fermion formalism [see, for
instance, Ref.~\onlinecite{Lee89}].

\section{Conclusion}

We have reconsidered the strong-coupling expansion for the Hubbard model
introduced by Sarker\cite{Sarker88} and Pairault {\it et
al.}\cite{Pairault98} The main modification is twofold. (i) We
introduce two different fermionic fields, $\gamma_+$ and
$\gamma_-$, which correspond to particles in the UHB and LHB,
respectively (Sec.~\ref{sec:sce}). (ii) We organize the
strong-coupling expansion around a ``non-degenerate'' ground-state where
each singly-occupied atomic state has a well defined spin direction
(which may fluctuate in time) (Sec.~\ref{sec:nond}). The action of the system,
$S[\gamma^*,\gamma;{\bf\Omega}]$, is then expressed in terms of two
fermionic fields, $\gamma_{-\uparrow}$ and $\gamma_{+\downarrow}$, and
a unit vector field ${\bf\Omega}_{\bf r}$. $\gamma_{-\uparrow}$ and
$\gamma_{+\downarrow}$ correspond to particles in the LHB and UHB,
respectively, the local spin-quantization axis being specified by the
time-fluctuating unit vector field ${\bf\Omega}_{\bf r}$. As in
Ref.~\onlinecite{Pairault98}, the strong-coupling expansion is carried
out by introducing a fermionic auxiliary field
$\psi\equiv(\psi_{-\uparrow},\psi_{+\downarrow})^T$ and performing a
Grassmannian Hubbard-Stratonovich transformation of the intersite
hopping term.  
 
Compared to the formalism discussed in Ref.~\onlinecite{Pairault98},
our work brings two major improvements. (i) The expansion involves a
single dimensionless parameter, $t/U$, and therefore does not break
down at low temperature. In the limit $U\to\infty$, the action
$S[\gamma^*,\gamma;{\bf\Omega}]$ becomes Gaussian
(Sec.~\ref{sec:Uinf}). At finite $U$, the corrections to the Gaussian
action can be obtained perturbatively by considering a finite number
of vertices. (ii) By introducing a fluctuating spin-quantization axis,
we have effectively suppressed the huge spin degeneracy of the atomic
ground-state. Indeed, for a given configuration of ${\bf\Omega}_{\bf
r}$, we now expand around a ``non-degenerate'' atomic state, where each
singly-occupied site carries a well defined spin (${\bf\Omega}_{\bf
r}/2$). 

This formalism allowed us to recover the action of the $t$-$J$ model
in the spin-hole coherent-state path integral.\cite{Auerbach91,Auerbach}
It can also be used to directly study the Hubbard model. It is
clear that the main technical difficulty comes from the spin variables
${\bf\Omega}_{\bf r}$, since the fermionic fields $\gamma_{-\uparrow}$
and $\gamma_{+\downarrow}$ can be integrated out in a systematic $t/U$
expansion. Thus further progress is tied to an
approximate treatment of spin fluctuations. The simplest approach
consists in expanding around a broken-symmetry ground-state by making
a saddle-point approximation on the spin variables ${\bf\Omega}_{\bf
r}$. In the half-filled Hubbard model, a natural choice is
${\bf\Omega}_{\bf r}=(-1)^{\bf r}\hat{\bf z}$, which corresponds to an AF
ground-state. Other choices, such as ferromagnetic or spiral orders,
are also possible. Work along these lines will be reported
elsewhere. \cite{ND00}

\section*{Acknowledgment} 

N.D. wishes to thank the CRPS and the Department of Physics of the
university of Sherbrooke for hospitality.

\bleq

\appendix

\section{}
\label{appI}

In this Appendix, we show how we can obtain the atomic Green's
functions {\it projected} on empty, singly occupied, and doubly
occupied sites. One could directly use the slave boson formulation by
explicitly integrating over the fields $f,e,p,d$ and the Lagrange
multipliers $\lambda$. This is the method used in Appendix
\ref{appIII}. Here we present a different (and more direct)
approach. We consider a single site and drop the site index. 

In the basis $\lbrace |0\rangle, |\uparrow\rangle, |\downarrow\rangle,
|\uparrow\downarrow \rangle \rbrace $, the operators
$\hat\gamma_{\alpha\sigma}$ are given by (in matrix form):
\begin{eqnarray}
\hat\gamma_{+\uparrow} = 
\left( \begin{array}{cccc}
0 & 0 & 0 & 0 \\ 0 & 0 & 0 & 0 \\ 0 & 0 & 0 & 1 \\ 0 & 0 & 0 & 0 
\end{array} \right ) ,
& \,\,\,\, &
\hat\gamma_{+\downarrow} = 
\left( \begin{array}{cccc}
0 & 0 & 0 & 0 \\ 0 & 0 & 0 & -1 \\ 0 & 0 & 0 & 0 \\ 0 & 0 & 0 & 0 
\end{array} \right ) ,  \nonumber \\ 
\hat\gamma_{-\uparrow} = 
\left( \begin{array}{cccc}
0 & 1 & 0 & 0 \\ 0 & 0 & 0 & 0 \\ 0 & 0 & 0 & 0 \\ 0 & 0 & 0 & 0 
\end{array} \right ) ,
&  \,\,\,\, &
\hat\gamma_{-\downarrow} = 
\left( \begin{array}{cccc}
0 & 0 & 1 & 0 \\ 0 & 0 & 0 & 0 \\ 0 & 0 & 0 & 0 \\ 0 & 0 & 0 & 0 
\end{array} \right ) . 
\label{A1}
\end{eqnarray}
From the equations of motion, we then deduce
\begin{eqnarray}
\hat\gamma_{+\sigma}(\tau) = e^{(\mu-U)\tau} \hat\gamma_{+\sigma} ,
&\,\,\,\,\,\,\,&
\hat\gamma^\dagger_{+\sigma}(\tau) =
e^{-(\mu-U)\tau}\hat\gamma^\dagger_{+\sigma} ,   \nonumber \\ 
\hat\gamma_{-\sigma}(\tau) = e^{\mu\tau} \hat\gamma_{-\sigma} , 
&\,\,\,\,\,\,\,&
\hat\gamma^\dagger_{-\sigma}(\tau) =
e^{-\mu\tau}\hat\gamma^\dagger_{-\sigma} ,
\end{eqnarray}
where $\hat\gamma^{(\dagger)}_{\alpha\sigma}(\tau)=
U(-\tau)\hat\gamma^{(\dagger)}_{\alpha\sigma} U(\tau)$, and 
\begin{equation}
U(\tau) = e^{-\tau(H-\mu N)} = 
\left( \begin{array}{cccc}
1 & 0 & 0 & 0 \\ 0 & e^{\mu\tau} & 0 & 0 \\ 0 & 0 & e^{\mu\tau} & 0 
\\ 0 & 0 & 0 & e^{(2\mu-U)\tau}
\end{array} \right ) .
\label{A3}
\end{equation}

Using (\ref{A1}) and (\ref{A3}), it is straightforward to calculate
the Green's functions. For the single-particle Green's function, we
find (for $\tau>0$) 
\begin{eqnarray}
G_{-\sigma}(\tau) &=& -\frac{1}{z} {\rm Tr}[U(\beta)
\hat\gamma_{-\sigma}(\tau) \hat\gamma^\dagger_{-\sigma}(0)] =
-\frac{e^{\mu\tau}}{z}  , \nonumber \\ 
G_{+\sigma}(\tau) &=& -\frac{1}{z} {\rm Tr}[U(\beta)
\hat\gamma_{+\sigma}(\tau) \hat\gamma^\dagger_{+\sigma}(0)] =
-\frac{e^{(\mu-U)\tau+\beta\mu}}{z} ,
\end{eqnarray}
where
\begin{equation}
z = {\rm Tr}\, U(\beta) = 1+2e^{\beta\mu}+e^{\beta(2\mu-U)}
\end{equation}
is the partition function of a single site. In Fourier space, we
obtain
\begin{eqnarray}
G_{-\sigma}(i\omega) &=& \frac{1-n_{\bar\sigma}}{i\omega+\mu} ,
\nonumber \\ 
G_{+\sigma}(i\omega) &=& \frac{n_{\bar\sigma}}{i\omega+\mu-U} ,
\end{eqnarray}
where 
\begin{equation}
n_\sigma = \frac{e^{\beta\mu}+e^{\beta(2\mu-U)}}{z} 
\end{equation}
is the mean number of $\sigma$-spin particles per site. The method is
straightforwardly extended to the calculation of higher-order Green's
functions.

\section{}
\label{appII}

In this Appendix we derive the spin-rotation-invariant action given by
Eqs.~(\ref{Stot}) and (\ref{Sat1}). 

In order to derive the functional integral, we introduce the mixed
fermion-boson coherent states $|\zeta\rangle \equiv |f,e,p,d\rangle$ defined by
\begin{equation}
|\zeta\rangle
= \exp \Bigl \lbrace \sum_{\bf r}\Bigl [ - \sum_\sigma 
f_{{\bf r}\sigma} \hat f^\dagger_{{\bf r}\sigma} + 
e_{\bf r} \hat e^\dagger_{\bf r} + 
\sum_\sigma p_{{\bf r}\sigma} \hat p^\dagger_{{\bf r}\sigma} +
d_{\bf r} \hat d^\dagger_{\bf r} \Bigr ] \Bigr \rbrace 
|{\rm vac}\rangle ,
\label{cs0}
\end{equation}
where $f_{{\bf r}\sigma}$ is a Grassmann variable, and $e_{\bf r}$,
$p_{{\bf r}\sigma}$ and $d_{\bf r}$ are c-numbers. In Eq.~(\ref{cs0}),
the spin-quantization axis is determined by the unit vector
${\bf\Omega}_{\bf r}$. The coherent states
$|\zeta\rangle$ satisfy the closure relation ($I$ is the unit operator) 
\begin{equation}
\int d\zeta^*d\zeta\, e^{-|\zeta|^2}|\zeta\rangle
\langle\zeta | P = I ,
\label{clos}
\end{equation}
where 
\begin{eqnarray}
|\zeta|^2 &=& \sum_{\bf r} \Bigl \lbrace \sum_\sigma 
f^*_{{\bf r}\sigma} f_{{\bf r}\sigma} + 
e^*_{\bf r} e_{\bf r} +
\sum_\sigma p^*_{{\bf r}\sigma} p_{{\bf r}\sigma} + 
d^*_{\bf r}  d_{\bf r} \Bigr \rbrace , \nonumber \\ 
\int d\zeta^*d\zeta &=& {\cal N}\int \prod_{\bf r}\bigl [ 
de^*_{\bf r} de_{\bf r}  dd^*_{\bf r} dd_{\bf r} 
\prod_\sigma (df^*_{{\bf r}\sigma} df_{{\bf r}\sigma}
dp^*_{{\bf r}\sigma} dp_{{\bf r}\sigma} ) \bigr ] , 
\end{eqnarray}
and $\cal N$ is a normalization constant. 
\begin{equation}
P=\prod_{\bf r}\Bigl( \delta_{Q_{\bf r}^{(1)},0} \prod_\sigma 
\delta_{Q_{{\bf r}\sigma}^{(2)},0} \Bigr)
\end{equation}
is a projection operator which ensures that the closure relation acts
only in the physical Hilbert space. The partition function is written as
\begin{equation}
Z= \int d\zeta^*d\zeta \, e^{-|\zeta|^2}
\langle \tilde \zeta |Pe^{-\beta(H -\mu N)}| \zeta \rangle , 
\label{Zapp1}
\end{equation}
where $\langle \tilde \zeta |=\langle -f,e,p,d|$. Splitting $\beta$
into $M$ intervals of width $\epsilon=\beta/M$ as in
Eq.~(\ref{Z2bis}), and introducing (M-1) times the closure relation
(\ref{clos}), we obtain
\begin{equation}
Z = \int \prod_{k=1}^M d\zeta^*_kd\zeta_k \,
e^{-\sum_{k=1}^M|\zeta_k|^2} \prod_{k=1}^M 
\langle  \zeta_k |P_k e^{-\epsilon(H -\mu N)}| \zeta_{k-1} \rangle ,
\end{equation}
with the boundary conditions $f_{{\bf r}\sigma,M}=-f_{{\bf r}\sigma,0}$, 
$e_{{\bf r},M}=e_{{\bf r},0}$, 
$p_{{\bf r}\sigma,M}=p_{{\bf r}\sigma,0}$, and 
$d_{{\bf r},M}=d_{{\bf r},0}$. The spin-quantization axis
${\bf\Omega}_{{\bf r},k}$ at a given site, defined by $R_{{\bf r},
k}\sigma_z R^\dagger_{{\bf r},k}=\bbox{\sigma}\cdot
{\bf\Omega}_{{\bf r},k}$, may depend on the ``time'' $k$ (but
satisfies the boundary condition ${\bf\Omega}_{{\bf
r},M}={\bf\Omega}_{{\bf r},0}$). The projection operator $P_k$ can be
written as
\begin{equation}
P_k=\prod_{\bf r} \Bigl( \int_0^{\frac{2\pi}{\epsilon}} \frac{\epsilon
d\lambda^{(1)}_{{\bf r},k}}{2\pi} e^{-i\epsilon\lambda^{(1)}_{{\bf
r},k} Q^{(1)}_{{\bf r},k}} 
\prod_\sigma \int_0^{\frac{2\pi}{\epsilon}}
\frac{\epsilon d\lambda^{(2)}_{{\bf r}\sigma,k}}{2\pi} 
e^{-i\epsilon\lambda^{(2)}_{{\bf r}\sigma,k} Q^{(2)}_{{\bf r}\sigma,k}} 
\Bigr ) .
\end{equation}
Note that the operators $P_k$, $Q^{(1)}_{{\bf r},k}$, and
$Q^{(2)}_{{\bf r}\sigma,k}$ are defined with respect to the
spin-quantization axis ${\bf\Omega}_{{\bf r},k}$. Since $Q^{(1)}_{{\bf
r},k}$ and $Q^{(2)}_{{\bf r}\sigma,k}$ commute with the Hamiltonian
$H-\mu N$, we have
\begin{eqnarray}
Z &=& \int \prod_{k=1}^M d\zeta^*_kd\zeta_k d\lambda_k\,
e^{-\sum_{k=1}^M|\zeta_k|^2} \prod_{k=1}^M 
\langle \zeta_k |e^{-\epsilon K_k}| \zeta_{k-1} \rangle , \nonumber \\ 
&=& \int \prod_{k=1}^M d\zeta^*_kd\zeta_k d\lambda_k \,
e^{-\sum_{k=1}^M|\zeta_k|^2} \prod_{k=1}^M \Bigl \lbrace 
\langle  \zeta_k | \zeta_{k-1} \rangle e^{-\epsilon 
K_k(\zeta^*_k,\zeta_{k-1})} \Bigr \rbrace 
\end{eqnarray}
in the limit $\epsilon\to 0$. We use the notations 
\begin{eqnarray}
\int d\lambda_k &=& \int \prod_{\bf r} \Bigl( \frac{\epsilon
d\lambda^{(1)}_{{\bf r},k}}{2\pi} \prod_\sigma
\frac{\epsilon d\lambda^{(2)}_{{\bf
r}\sigma,k}}{2\pi} \Bigr) , \nonumber \\ 
K_k &=& H-\mu N +i\sum_{\bf r}(\lambda^{(1)}_{{\bf r},k}Q^{(1)}_{{\bf
r},k}+\sum_\sigma \lambda^{(2)}_{{\bf r}\sigma,k} Q^{(2)}_{{\bf
r}\sigma,k}) ,   \nonumber \\ 
K_k(\zeta^*_k,\zeta_{k-1}) &=& \frac{\langle\zeta_k | K_k |
\zeta_{k-1} \rangle } {\langle\zeta_k|\zeta_{k-1} \rangle } .
\end{eqnarray}

The only difference with the standard derivation of the functional
integral\cite{Negele}  for a mixed fermion-boson system comes from the
evaluation of the 
scalar product $\langle  \zeta_k | \zeta_{k-1} \rangle$. 
If $R_{{\bf r},k}= R_{{\bf r},k-1}$ at each lattice site, we have the
usual result  
\begin{equation}
\langle  \zeta_k | \zeta_{k-1} \rangle = \exp \Bigl \lbrace 
\sum_r \Bigl [ \sum_\sigma 
f^*_{{\bf r}\sigma,k} f_{{\bf r}\sigma,k-1} + 
e^*_{{\bf r},k} e_{{\bf r},k-1}  + 
\sum_\sigma p^*_{{\bf r}\sigma,k} p_{{\bf r}\sigma,k-1}
+ d^*_{{\bf r},k} d_{{\bf r},k-1} 
\Bigr ] \Bigl \rbrace ,
\label{psca}
\end{equation}
since the spin-quantization is the same at ``times'' $k$ and $k-1$. In the
more general situation where the spin-quantization may fluctuate in
time, we rewrite the coherent state $|\zeta_k \rangle$ as
\begin{eqnarray}
|\zeta_k \rangle &=& \prod_{\bf r}  \bigl \lbrace 
e_{{\bf r},k} \hat e^\dagger_{{\bf r},k} - \sum_\sigma 
f_{{\bf r}\sigma,k} \hat f^\dagger_{{\bf r}\sigma,k}
p_{{\bf r}\sigma,k} \hat p^\dagger_{{\bf r}\sigma,k} - 
f_{{\bf r}\uparrow,k} f_{{\bf r}\downarrow,k}
\hat f^\dagger_{{\bf r}\uparrow,k} \hat f^\dagger_{{\bf r}\downarrow,k}
d_{{\bf r},k} \hat d^\dagger_{{\bf r},k} \bigr \rbrace |{\rm
vac}\rangle  \nonumber \\ &=& 
\prod_{\bf r}  \bigl \lbrace 
e_{{\bf r},k} |0,{\bf r}\rangle -  \sum_\sigma 
f_{{\bf r}\sigma,k} p_{{\bf r}\sigma,k} |\sigma,{\bf r},k\rangle - 
f_{{\bf r}\uparrow,k} f_{{\bf r}\downarrow,k} d_{{\bf r},k}
|\uparrow\downarrow,{\bf r}\rangle \bigr \rbrace,
\label{coh1}
\end{eqnarray}
using the constraints (\ref{con}). $ |0,{\bf r}\rangle$, $|\sigma,{\bf
r},k\rangle$, and $|\uparrow\downarrow,{\bf r}\rangle$ denote the
empty, singly occupied, and doubly occupied states, respectively [see
Eq.~(\ref{atst})]. The singly occupied state $|\sigma,{\bf r},k\rangle$
depends on the ``time'' $k$ since the particle has its spin quantized
along the ${\bf\Omega}_{{\bf r},k}$-axis. From Eq.~(\ref{coh1}), we
deduce 
\begin{eqnarray}
\langle \zeta_k|\zeta_{k-1} \rangle &=& \prod_{\bf r} \bigl \lbrace e^*_{{\bf
r},k}e_{{\bf r},k-1} + \sum_{\sigma,\sigma'} f^*_{{\bf r}\sigma,k}
p^*_{{\bf r}\sigma,k}  f_{{\bf r}\sigma',k-1} p_{{\bf r}\sigma',k-1}
\langle \sigma,{\bf r},k| \sigma',{\bf r},k-1 \rangle  \nonumber \\ &&
+ f^*_{{\bf r}\downarrow,k} f^*_{{\bf r}\uparrow,k}
f_{{\bf r}\uparrow,k-1} f_{{\bf r}\downarrow,k-1}
d^*_{{\bf r},k} d_{{\bf r},k-1} \bigr \rbrace .
\label{psca2}
\end{eqnarray}
When ${\bf\Omega}_{{\bf r},k}={\bf\Omega}_{{\bf r},k-1}$, $\langle
\sigma,{\bf r},k| \sigma',{\bf r},k-1 \rangle =
\delta_{\sigma,\sigma'}$. Eq.~(\ref{psca2}) is then equivalent to
Eq.~(\ref{psca}). [This equivalence is due to the constraints
(\ref{con}).] For ${\bf\Omega}_{{\bf r},k}\simeq {\bf\Omega}_{{\bf
r},k-1}$, this allows to rewrite Eq.~(\ref{psca2}) as
\begin{eqnarray}
\langle \zeta_k|\zeta_{k-1} \rangle &=& \exp \Bigl \lbrace 
\sum_r \Bigl [ \sum_\sigma 
f^*_{{\bf r}\sigma,k} f_{{\bf r}\sigma,k-1} + 
e^*_{{\bf r},k} e_{{\bf r},k-1}  + 
\sum_\sigma p^*_{{\bf r}\sigma,k} p_{{\bf r}\sigma,k-1}
+ d^*_{{\bf r},k} d_{{\bf r},k-1} \nonumber \\ &&
+ \sum_{\sigma,\sigma'} f^*_{{\bf r}\sigma,k}
p^*_{{\bf r}\sigma,k}  f_{{\bf r}\sigma',k-1} p_{{\bf r}\sigma',k-1}
\bigl ( (R^\dagger_{{\bf r},k}R_{{\bf r},k-1})_{\sigma\sigma'}
-\delta_{\sigma,\sigma'} \bigr ) \Bigr ] \Bigl \rbrace ,
\end{eqnarray}
where we have used $\langle \sigma,{\bf r},k| \sigma',{\bf r},k-1 \rangle =
(R^\dagger_{{\bf r},k}R_{{\bf r},k-1})_{\sigma\sigma'}$. 
Thus we may write the partition function $Z$ as
\begin{eqnarray}
Z &=& \int \prod_{k=1}^M d\zeta^*_kd\zeta_k d\lambda_k\, \exp \Bigl
\lbrace -\sum_{k=1}^M \sum_{\bf r} \Bigl [ \sum_\sigma 
f^*_{{\bf r}\sigma,k} (f_{{\bf r}\sigma,k}-f_{{\bf r}\sigma,k-1})
+ e^*_{{\bf r},k} (e_{{\bf r},k}-e_{{\bf r},k-1})
\nonumber \\ && 
+\sum_\sigma p^*_{{\bf r}\sigma,k} (p_{{\bf r}\sigma,k}-p_{{\bf r}\sigma,k-1})
+ d^*_{{\bf r},k} (d_{{\bf r},k}-d_{{\bf r},k-1})  \nonumber \\ && 
- \sum_{\sigma,\sigma'}  f^*_{{\bf r}\sigma,k}
p^*_{{\bf r}\sigma,k}  f_{{\bf r}\sigma',k-1} p_{{\bf r}\sigma',k-1}
\bigl ( (R^\dagger_{{\bf r},k}R_{{\bf r},k-1})_{\sigma\sigma'}
-\delta_{\sigma,\sigma'} \bigr ) \Bigr ]
- \epsilon \sum_{k=1}^M K_k(\zeta^*_k,\zeta_{k-1}) \Bigr \rbrace .
\end{eqnarray}
Spin-rotation invariance is obtained by summing over all possible
configurations of the unit vector ${\bf\Omega}_{{\bf r},k}$, i.e.
\begin{equation}
Z \to \int \prod_{\bf r} \prod_{k=1}^M 
\frac{d{\bf\Omega}_{{\bf r},k}}{4\pi} Z .
\end{equation}
When taking the continuum limit, one has to evaluate
$K_k(\zeta^*_k,\zeta_k)$ (instead of
$K_k(\zeta^*_k,\zeta_{k-1})$). Since all the operators involved in that
quantity are defined with respect to the same spin-quantization axis
(one can choose the axis ${\bf\Omega}_{{\bf r},k}$ to express the
Hamiltonian $H-\mu N$), $K_k(\zeta^*_k,\zeta_k)$ is readily
calculated. We finally obtain 
\begin{equation}
Z = \int {\cal D}{\bf\Omega} \int {\cal D}{\lambda} 
\int {\cal D}[f,e,p,d]\, e^{-S},  
\label{Zll}
\end{equation}
where the action $S$ is given by 
\begin{eqnarray}
S &=& S_{\rm at} -\sum_{{\bf r},{\bf r}',\alpha,\alpha'} \int d\tau \, 
\gamma^\dagger_{{\bf r}\alpha} R^\dagger_{\bf r}t_{{\bf rr}'}R_{{\bf
r}'} \gamma_{{\bf r}'\alpha'} , \label{Sat0a} \\ 
S_{\rm at} &=& S_{\rm at}^{(0)} + \sum_{{\bf r},\sigma,\sigma'} \int
d\tau \, f^*_{{\bf r}\sigma}p^*_{{\bf r}\sigma} (R^\dagger_{\bf r}\dot R_{\bf
r})_{\sigma\sigma'} f_{{\bf r}\sigma'}p_{{\bf r}\sigma'} , 
\label{Sat0b}  \\ 
S_{\rm at}^{(0)} &=& \sum_{\bf r}
\int d\tau\, \Bigl [ -i\lambda^{(1)}_{\bf r}+\sum_\sigma  f^*_{{\bf r}\sigma}
(\partial_\tau-\mu+i\lambda^{(2)}_{{\bf r}\sigma}) f_{{\bf r}\sigma} + 
e^*_{\bf r} (\partial_\tau+i\lambda^{(1)}_{\bf r}) e_{\bf r}
\nonumber \\ && + \sum_\sigma  p^*_{{\bf r}\sigma}
(\partial_\tau+i\lambda^{(1)}_{\bf r}-i\lambda^{(2)}_{{\bf r}\sigma})
p_{{\bf r}\sigma}
+ d^*_{\bf r} (\partial_\tau+i\lambda^{(1)}_{\bf
r}-i\sum_\sigma\lambda^{(2)}_{{\bf r}\sigma}+U) d_{\bf r} \Bigr ] .
\label{Sat0}
\end{eqnarray}
The variables $\gamma$ are defined in Sec.~\ref{sec:nond}. It is well
known that in the standard slave boson approach, the Lagrange
multipliers can be chosen to be time independent since the constraints are
preserved under time evolution.\cite{Kotliar86} The action $S$
[Eqs.~(\ref{Sat0a}-\ref{Sat0})]  
differs from the standard action by the additional term proportional
to $R^\dagger\dot R$ which comes from time fluctuations of the
spin-quantization axis. This term conserves the total
number of $p$ bosons and therefore the constraints $Q^{(1)}_{\bf r}$. It is
also clear that it lets the combination $f^*_{{\bf r}\sigma}f_{{\bf
r}\sigma} - p^*_{{\bf r}\sigma}p_{{\bf r}\sigma}$ invariant, so that
the constraints $Q^{(2)}_{{\bf r}\sigma}$ are also conserved. Consequently, we
can replace the functional integral over $\lambda(\tau)$ in
Eq.~(\ref{Zll}) by an integral over a set $\lambda\equiv (\lambda^
{(1)}_{\bf r},\lambda^{(2)}_{{\bf r}\sigma})$ of time-independent
Lagrange multipliers:
\begin{equation}
\int {\cal D}\lambda \to \int d\lambda \equiv \prod_{\bf r} \Bigl
( \int_0^\frac{2\pi}{\beta} \frac{\beta d\lambda^{(1)}_{\bf r}}{2\pi}
\prod_\sigma  \int_0^\frac{2\pi}{\beta} \frac{\beta
d\lambda^{(2)}_{{\bf r}\sigma}}{2\pi} \Bigr ) .
\end{equation}
Note that the introduction of time-dependent
Lagrange multipliers at an intermediate stage is what allows a simple
evaluation of $K_k(\zeta^*_k,\zeta_{k-1})$. To avoid this
complication, one would have to consider the $p$ operators as $2\times
2$ matrices in spin space.

\section{}
\label{appIII}

In this Appendix, we calculate the single-particle
Green's functions $G^{(0)}_\sigma(\tau)$ and $G_\sigma(\tau)$
corresponding to the action $S^{(0)}_{\rm at}$ and $S_{\rm at}$, respectively
[Eqs.~(\ref{Sgam})]. 

$G^{(0)}_\sigma(\tau)$ and $G_\sigma(\tau)$ are
obtained directly from their expressions in terms of the fields
$f_\sigma,e,p_\uparrow$ and $d$. As discussed in Sec.~\ref{sec:nond},
the chemical 
potential $\mu$ equals $U/2$ at half-filling and lies near the top of
the LHB in the presence of a finite concentration of holes. The
ground-state of the system in the atomic limit then has exactly one
particle per site (even away from half-filling).  We consider a
single site and drop the site index.

Let us first consider $G^{(0)}_\sigma(\tau)$. It can be written as
\begin{eqnarray}
G^{(0)}_\uparrow(\tau) &=& -\frac{1}{z^{(0)}} \int d\lambda \int {\cal
D}[f,e,p,d]\,e^*(\tau)p_\uparrow(\tau)f_\uparrow(\tau)
f^*_\uparrow(0)p^*_\uparrow(0)e(0)\exp \bigl \lbrace -S^{(0)}_{\rm at}
\bigr \rbrace \nonumber \\ &=& -\frac{1}{z^{(0)}} \int d\lambda \,
z^{(0)}_\lambda 
g^{(0)}_{\uparrow}(\tau) , \nonumber \\ 
G^{(0)}_{\downarrow}(\tau) &=& -\frac{1}{z^{(0)}} \int d\lambda \int {\cal
D}[f,e,p,d]\,d(\tau)p^*_{\uparrow}(\tau)f_\downarrow(\tau)
f^*_\downarrow(0)p_{\uparrow}(0)d^*(0)\exp \bigl \lbrace -S^{(0)}_{\rm at}
\bigr \rbrace \nonumber \\ &=& -\frac{1}{z^{(0)}} \int d\lambda \,
z^{(0)}_\lambda g^{(0)}_{\downarrow}(\tau) ,
\label{C1}
\end{eqnarray}
where 
\begin{eqnarray}
z^{(0)}_\lambda &=&  \int {\cal
D}[f,e,p,d]\, \exp \bigl \lbrace -S^{(0)}_{\rm at} \bigr \rbrace , \nonumber \\
g^{(0)}_{\uparrow}(\tau) &=& -\frac{1}{z^{(0)}_\lambda}
\int {\cal D}[f,e,p,d]\,e^*(\tau)p_\uparrow(\tau)f_\uparrow(\tau)
f^*_\uparrow(0)p^*_\uparrow(0)e(0)\exp \bigl \lbrace -S^{(0)}_{\rm at}
\bigr \rbrace  , \nonumber \\ 
g^{(0)}_{\downarrow}(\tau) &=& -\frac{1}{z^{(0)}_\lambda}
\int {\cal D}[f,e,p,d]\,d(\tau)p^*_{\uparrow}(\tau)f_\downarrow(\tau)
f^*_\downarrow(0)p_{\uparrow}(0)d^*(0)\exp \bigl \lbrace -S^{(0)}_{\rm at}
\bigr \rbrace
\end{eqnarray}
are the partition function and the Green's functions for a given
configuration of the Lagrange multipliers. $z^{(0)}=\int
d\lambda\,z^{(0)}_\lambda$ is the partition function for a single
site. Since $S^{(0)}_{\rm at}$ is a Gaussian action when $\lambda$ is
held fixed, we immediately deduce 
\begin{eqnarray}
g^{(0)}_{\uparrow}(\tau) &=&
g_{f_\uparrow}(\tau)g_{p_\uparrow}(\tau)g_e(-\tau), 
\nonumber \\ 
g^{(0)}_{\downarrow}(\tau) &=&
g_{f_\downarrow}(\tau)g_{p_{\uparrow}}(-\tau) g_d(\tau),
\end{eqnarray}
where 
\begin{eqnarray}
g_{f_\sigma}(\tau) &=&  -e^{(\mu-i\lambda^{(2)}_\sigma)\tau}
\bigl [ \theta(\tau)(1-F_\sigma)-\theta(-\tau)F_\sigma \bigr ] ,
\nonumber \\ 
g_e(\tau) &=& -e^{-i\lambda^{(1)}\tau}
\bigl [ \theta(\tau)(1+B)+\theta(-\tau)B \bigr ] ,
\nonumber \\ 
g_{p_\uparrow}(\tau) &=& -e^{-(i\lambda^{(1)}-i\lambda^{(2)}_\uparrow)\tau}
\bigl [ \theta(\tau)(1+B_\uparrow)+\theta(-\tau)B_\uparrow \bigr ] ,
\nonumber \\ 
g_d(\tau) &=& -e^{-(i\lambda^{(1)}-i\sum_\sigma \lambda^{(2)}_\sigma+U)\tau}
\bigl [ \theta(\tau)(1+B')+\theta(-\tau)B' \bigr ] ,
\label{C4}
\end{eqnarray}
are the propagators of the fields $f_\sigma$, $e$, $p_\uparrow$ and $d$ when
the Lagrange multipliers are held fixed. We use the notation
\begin{eqnarray}
F_\sigma &=& n_F(-\mu+i\lambda_\sigma^{(2)}) , \nonumber \\ 
B_\uparrow &=& n_B(i\lambda^{(1)}-i\lambda_\uparrow^{(2)}) , \nonumber \\ 
B &=& n_B(i\lambda^{(1)}) , \nonumber \\ 
B' &=& n_B(i\lambda^{(1)}-i\sum_\sigma \lambda^{(2)}_\sigma+U) ,
\end{eqnarray}
where $n_F(x)=(e^{\beta x}+1)^{-1}$ and $n_B(x)=(e^{\beta x}-1)^{-1}$
are the Fermi and Bose factors, respectively. Using  
\begin{equation}
z^{(0)}_\lambda = \prod_\sigma
\bigl(1+e^{\beta(\mu-i\lambda^{(2)}_\sigma)} \bigr) 
\bigl( e^{i\beta\lambda^{(1)}}+1+ e^{i\beta\lambda^{(2)}_\uparrow} +
e^{-\beta(U-i\sum_\sigma\lambda^{(2)}_\sigma)} \bigr)
+O(e^{-i\beta\lambda^{(1)}}) , 
\end{equation}
we obtain
\begin{equation}
z^{(0)} = \int d\lambda\, z^{(0)}_\lambda = 1+e^{\beta\mu}+
e^{\beta(2\mu-U)} \simeq  e^{\beta\mu}
\end{equation}
in the limit $T\to 0$. In order to perform the integration over the
Lagrange multipliers, we note that 
\begin{eqnarray}
B &=& e^{-i\beta\lambda^{(1)}} + O(e^{-2i\beta\lambda^{(1)}}) , \nonumber \\ 
B' &=& e^{-i\beta\lambda^{(1)}+i\beta\sum_\sigma\lambda^{(2)}_\sigma-\beta U} 
+ O(e^{-2i\beta\lambda^{(1)}}) , \nonumber \\ 
B_\uparrow &=& e^{-i\beta\lambda^{(1)}+i\beta\lambda^{(2)}_\uparrow} +
O(e^{-2i\beta\lambda^{(1)}}) .
\label{C8}
\end{eqnarray}
We deduce from Eqs.~(\ref{C1}), (\ref{C4}) and (\ref{C8})
\begin{eqnarray}
G^{(0)}_{\uparrow}(\tau) &=& - \frac{e^{\mu\tau}}{z^{(0)}} \bigl [ 
\theta(\tau)-\theta(-\tau)e^{\beta\mu} \bigr ] \to
\theta(-\tau+\eta) e^{\mu\tau} , \nonumber \\ 
G^{(0)}_{\downarrow}(\tau) &=&  - \frac{e^{(\mu-U)\tau}}{z^{(0)}} \bigl [ 
\theta(\tau)e^{\beta\mu}-\theta(-\tau)e^{\beta(2\mu-U)}\bigr ]
\to - \theta(\tau-\eta) e^{(\mu-U)\tau} ,
\label{CG}
\end{eqnarray}
where we have taken the limit $T\to 0$. The
infinitesimal $\eta$ ($\eta\to 0^+$) is introduced to properly define
the equal-time Green's functions. \cite{Negele}

Let us now consider the correction due to 
\begin{equation}
S_{\rm at}-S^{(0)}_{\rm at} = \int d\tau \,
A^0_z f^*_\uparrow p^*_\uparrow f_\uparrow p_\uparrow .
\end{equation}
Since $G_\sigma=z^{-1}\int d\lambda\, z_\lambda g_\sigma$,
the correction of order $O(A^0_z)$ is given by
\begin{equation}
G^{(1)}_\sigma = -
\frac{z^{(1)}}{z^{(0)}}G^{(0)}_\sigma 
+ \frac{1}{z^{(0)}}\int d\lambda\,z^{(1)}_\lambda g^{(0)}_\sigma   
+ \frac{1}{z^{(0)}}\int d\lambda\,z^{(0)}_\lambda g^{(1)}_\sigma,
\end{equation}
where $z^{(1)}_\lambda$ and $g^{(1)}_\sigma$ are the first-order
corrections to $z^{(0)}_\lambda$ and $g^{(0)}_\sigma$. The
correction to the partition function is given by
\begin{eqnarray}
z^{(1)}_\lambda &=& -z^{(0)}_\lambda  F_\uparrow B_\uparrow
\int d\tau \, A^0_z , \label{C20a} \\ 
z^{(1)} &=& \int d\lambda \, z^{(1)}_\lambda = 
- e^{\beta\mu} \int d\tau \, A^0_z .
\label{C20b}
\end{eqnarray}
From (\ref{C20a}), we deduce 
\begin{equation}
\int d\lambda \,z^{(1)}_\lambda g^{(0)}_{\sigma} = 0 . 
\label{C30}
\end{equation}

\begin{figure}
\epsfxsize 7.5cm 
\epsffile[58 340 300 728]{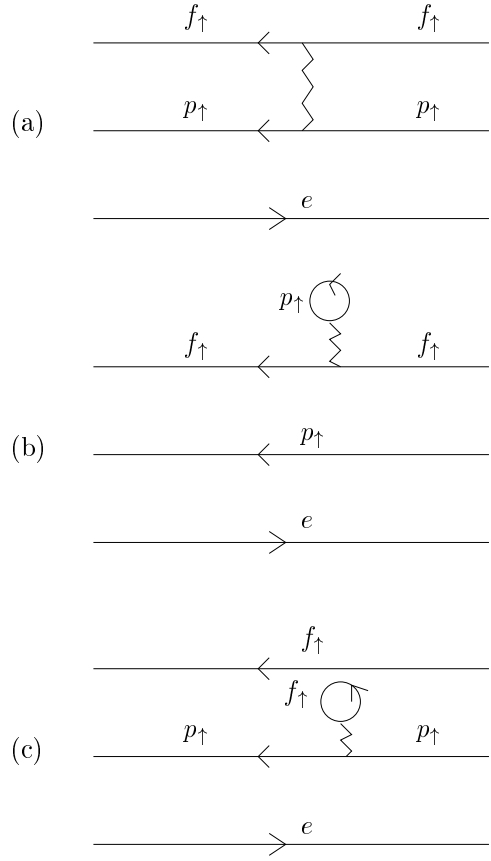}
\caption{
Diagrammatic representation of the first-order ($O(A^0_z)$) correction
$g^{(1)}_{\uparrow}$ to  $g^{(0)}_{\uparrow}$. The wavy line denotes
the gauge field $A^0_z$. 
}
\label{Fig2}
\end{figure}

\begin{figure}
\epsfxsize 7.5cm 
\epsffile[58 450 300 613]{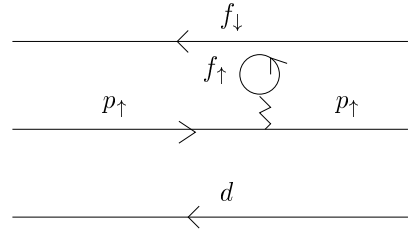}
\caption{Diagrammatic representation of the first-order  ($O(A^0_z)$)
correction $g^{(1)}_{\downarrow}$ to  $g^{(0)}_{\downarrow}$.}
\label{Fig3}
\end{figure}

The first-order corrections to $g^{(0)}_{\uparrow}$ and
$g^{(0)}_{\downarrow}$ are shown diagrammatically in
Figs.~\ref{Fig2} and \ref{Fig3}:
\begin{eqnarray}
g^{(1)}_{\uparrow}(\tau) &=& - \int d\tau_1 \, A^0_z(\tau_1) 
g_{f_\uparrow}(\tau-\tau_1) g_{f_\uparrow}(\tau_1) 
g_{p_\uparrow}(\tau-\tau_1) g_{p_\uparrow}(\tau_1) 
g_e(-\tau)  \nonumber \\  &&
- \int d\tau_1 \, A^0_z(\tau_1) 
g_{f_\uparrow}(\tau-\tau_1) g_{f_\uparrow}(\tau_1) 
g_{p_\uparrow}(0^-) g_{p_\uparrow}(\tau)  
g_e(-\tau)  \nonumber \\  &&
+ \int d\tau_1 \, A^0_z(\tau_1) g_{f_\uparrow}(\tau)
g_{p_\uparrow}(\tau-\tau_1) g_{p_\uparrow}(\tau_1) 
 g_{f_\uparrow}(0^-)
g_e(-\tau) , \nonumber \\
g^{(1)}_{\downarrow}(\tau) &=&
\int d\tau_1 \, A^0_z(\tau_1) g_{f_\downarrow}(\tau) 
g_{p_\uparrow}(-\tau+\tau_1)g_{f_\uparrow}(0^-)g_{p_\uparrow}(-\tau_1)
g_d(\tau) .
\end{eqnarray}
Integrating over the Lagrange multipliers, we obtain
\begin{eqnarray}
\int d\lambda\,z^{(0)}_\lambda g^{(1)}_{\uparrow}
&=& e^{\mu\tau} \int d\tau_1\, A^0_z(\tau_1)
\bigl \lbrace \theta(\tau)\theta(\tau-\tau_1)\theta(\tau_1)
\nonumber \\ && 
-e^{\beta\mu} \theta(-\tau)[\theta(\tau-\tau_1)\theta(-\tau_1) +
\theta(-\tau+\tau_1)\theta(\tau_1)] \bigr \rbrace ,  \nonumber \\ 
\int d\lambda\,z^{(0)}_\lambda g^{(1)}_{\downarrow}
&=& e^{(\mu-U)\tau} \int d\tau_1\, A^0_z(\tau_1)
\bigl \lbrace e^{\beta\mu} \theta(\tau) 
[\theta(\tau-\tau_1)\theta(-\tau_1) +
\theta(-\tau+\tau_1)\theta(\tau_1)] \nonumber \\ && 
-e^{\beta(2\mu-U)}\theta(-\tau)\theta(-\tau+\tau_1)\theta(-\tau_1)
\bigr \rbrace .
\label{C40}
\end{eqnarray}
From Eqs.~(\ref{C20b}), (\ref{C30}) and (\ref{C40}), we finally obtain
\begin{eqnarray}
G^{(1)}_{\uparrow} (\tau) &=& \int d\tau_1 \, 
G^{(0)}_{\uparrow} (\tau-\tau_1)A^0_z(\tau_1)G^{(0)}_{\uparrow}
(\tau_1) , \nonumber \\ 
G^{(1)}_{\downarrow} (\tau) &=& - \int d\tau_1 \, 
G^{(0)}_{\downarrow} (\tau-\tau_1)A^0_z(\tau_1)G^{(0)}_{\downarrow}
(\tau_1) 
\end{eqnarray}
in the limit $T\to 0$. 
These results are easily extrapolated to higher orders as (in matrix form)
\begin{eqnarray}
G_{\uparrow} &=& G^{(0)}_{\uparrow} + 
G^{(0)}_{\uparrow} A^0_z G^{(0)}_{\uparrow} + 
G^{(0)}_{\uparrow} A^0_z G^{(0)}_{\uparrow} A^0_z G^{(0)}_{\uparrow} 
+ \cdots \nonumber \\ 
G_{\downarrow} &=& G^{(0)}_{\downarrow} - 
G^{(0)}_{\downarrow} A^0_z G^{(0)}_{\downarrow} + 
G^{(0)}_{\downarrow} A^0_z G^{(0)}_{\downarrow} A^0_z G^{(0)}_{\downarrow} 
- \cdots 
\end{eqnarray}
or, equivalently,
\begin{eqnarray}
G^{-1}_{\uparrow} &=& G^{(0)-1}_{\uparrow} - A^0_z , \nonumber \\ 
G^{-1}_{\downarrow} &=& G^{(0)-1}_{\downarrow} + A^0_z .
\end{eqnarray}
Note that these Green's functions, like the gauge field $A^0_z$,
depend on the site which is considered.

\eleq


\newpage

\ecols


\begin{references}
\bibitem{note1} One then has to modify the measure in the functional
integral accordingly, in order to let the value of the partition
function unchanged.
\end{references}
\end{document}